\documentclass[12pt,a4paper]{article}
\usepackage{jheppub}

\usepackage[table,dvipsnames]{xcolor}
\usepackage{amsmath,latexsym,amssymb}
\usepackage{graphicx}
\usepackage{psfrag}
\usepackage[latin1]{inputenc}
\usepackage{subfigure}
\usepackage{nicefrac}
\usepackage{bbm}
\usepackage[stable]{footmisc}
\usepackage{rotating}
\usepackage{afterpage}
\usepackage{multirow}

\usepackage{comment}

\newcommand\nn{\nonumber}
\newcommand\bbone{\ensuremath{\mathbbm{1}}}

\newcommand{\eq}[1]{\begin{equation}#1\end{equation}}
\newcommand{\spl}[1]{\begin{split}#1\end{split}}
\newcommand{\al}[1]{\begin{align}#1\end{align}}
\newcommand{\subeq}[1]{\begin{subequations}#1\end{subequations}}

\def\d{\text{d}}

\def\Re           {{\rm Re\hskip0.1em}}
\def\Im           {{\rm Im\hskip0.1em}}

\newcommand{\be}{\begin{equation}}
\newcommand{\ee}{\end{equation}}
\newcommand{\bea}{\begin{eqnarray}}
\newcommand{\eea}{\end{eqnarray}}

\definecolor{propfield}{rgb}{0, 0.8, 0.2}

\title{Tri-Sasakian consistent reduction}

\author[\clubsuit]{Davide Cassani}
\author[\diamondsuit]{and Paul Koerber}

\affiliation[\clubsuit]{Department of Mathematics, King's College London,\\
The Strand, London WC2R 2LS, United Kingdom\\
}

\emailAdd{davide.cassani at kcl.ac.uk}

\affiliation[\diamondsuit]{Instituut voor Theoretische Fysica, Katholieke Universiteit Leuven\\
Celestijnenlaan 200D, B-3001 Leuven, Belgium}

\emailAdd{koerber at itf.fys.kuleuven.be}

\abstract{We establish a universal consistent Kaluza--Klein truncation of M-theory
based on seven-dimensional tri-Sasakian structure. The four-dimensional truncated theory is an $\mathcal{N}=4$ gauged
supergravity with three vector multiplets and a non-abelian gauge group, containing the compact factor SO(3). Consistency follows from the fact that our truncation takes exactly the same form as a left-invariant reduction on a specific coset manifold, and we show that the same holds for the various universal consistent truncations recently put forward in the literature.
We describe how the global symmetry group SL$(2,\mathbb{R})\times \text{SO}(6,3)$ is embedded in the symmetry group E$_{7(7)}$ of
maximally supersymmetric reductions, and make the connection with the approach of Exceptional
Generalized Geometry. Vacuum AdS$_4$ solutions spontaneously break the amount of supersymmetry from $\mathcal N=4$ to $\mathcal N=3,1$ or $0$, and the spectrum contains massive modes.
We find a subtruncation to minimal $\mathcal{N} = 3$ gauged supergravity as well as an $\mathcal{N} = 1$ subtruncation to the SO(3)-invariant sector.  We also show that a reduction on the homogeneous space $N^{010}$ enhances the universal tri-Sasakian truncation with a Betti vector multiplet.}

\begin{document}

\begin{flushright}   {\small KCL-MTH-11-19}
\end{flushright}
\vspace{-1.03cm}

\maketitle
\flushbottom

\section{Introduction and discussion of the results}

Consistent truncations provide a powerful method to construct solutions of a complicated theory
by uplifting solutions of a simpler, truncated theory. A truncation is consistent if every solution of the equations of motion
of the truncated theory can be lifted to a solution of the equations of motion
of the untruncated theory. This fails if the modes of the truncated theory source
modes that are truncated out, so that it is inconsistent to put the latter to zero while keeping the former. Typically, the original theory is
a higher-dimensional theory, like 10D and 11D supergravity, which is then compactified on some internal
manifold $M$ down to a lower dimension. The truncation then selects at the non-linear level a finite subset of the infinite tower of Kaluza--Klein (KK) modes.

A consistent truncation requires a delicate canceling of terms, such as in the prominent examples of maximally supersymmetric
reduction on spheres preserving all massless KK modes: 11D supergravity on $S^7$ \cite{deWitS7} or $S^4$ \cite{nasvam1,nasvam2}, and IIB supergravity on $S^5$ \cite{pilchwarnerS5,SO6truncationIIB}.
While for the former two cases consistency is established, for the last one only a partial proof has been given. Consistent truncations are therefore considered rare.
However, in the case there is a global symmetry group $G$ and one restricts to fields that are invariant under
$G$, consistency is guaranteed. Indeed, non-linear terms built from $G$-singlet modes cannot act as sources for the truncated
non-singlet modes. This is for instance the case for compactifications on a coset manifold $M=G/H$, where one restricts to
fields that are invariant under the action of $G$ (for recent examples see e.g.~\cite{ExploitingN=2}).

Recently, consistent truncations of a {\em universal} type, which hold for a whole class of compactification manifolds $M$,
have been constructed. Building on a few previous examples, in \cite{vargaun1} it was conjectured that for every supersymmetric warped AdS solution of 10D or 11D supergravity on a Riemannian manifold,
there is a consistent truncation on $M$ containing just the supergravity multiplet. It became more
interesting in \cite{vargaun2}, where a consistent truncation of 11D supergravity on a generic 7D Sasaki--Einstein
space was established. This time the truncated theory is richer than just minimal $\mathcal{N}=2$ supergravity, in that it
contains extra multiplets of massive fields, including a breathing mode (overall volume of $M$), a squashing mode (relative volume of
the local fiber and K\"ahler--Einstein base of the Sasaki--Einstein manifold) and more scalars from the form fields.
This work was preceded by \cite{maldtach}, and followed by \cite{IIBonSE,vargaun3,LiuSzepietowskiZhao,tsimpistaylor}, where new universal consistent truncations of type IIB supergravity based on 5D Sasaki--Einstein structures were constructed.
In \cite{T11red,granabena} the special case of $T^{1,1}$ was considered, which compared to the generic Sasaki--Einstein case
allows for an extra $\mathcal N=4$ vector multiplet associated to the non-trivial cohomology of $T^{1,1}$, which is therefore dubbed a {\em Betti multiplet} and which turns out to be crucial in the gauge-gravity applications based on conifold geometries. Although in the above papers just the bosonic sector of the reduction was studied, the arguments on which supersymmetry and consistency of the reduction rely also hold for the fermionic sector. This has been explicitly studied in \cite{Bah:2010yt,Bah:2010cu,Liu:2010pq,Liu:2011dw}.

Interest in these new consistent truncations was spurred by the fact that the truncated theories comprised both charged and massive fields, and could be used to embed solutions proposed to be dual to condensed matter systems into string or M-theory. For instance, holographic superconductors \cite{Gubser_superconductor,HHH1} and solutions with non-relativistic scale invariance of the Schr\"odinger \cite{Son_Schroedinger,BalaMcGreevy_Schroedinger} or Lifshitz \cite{kachrulifshitz} type were lifted using this approach, see e.g.~\cite{Denef:2009tp,SupercSupers,gauntlettholo1,Gubser:2009gp,gauntlettholo2,Aprile:2011uq,Donos:2011ut,Bobev:2011rv}, \cite{maldtach, vargaun2}, and \cite{gauntlettdonos,Gregory:2010gx,gauntlettdonos2,cassanilifshitz} respectively, as well as \cite{OColgainSamtleben} for further applications.
Moreover, these universal truncations also admit AdS backgrounds, and hence dual conformal points. As they represent a consistent subsector of the higher-dimensional theory which is common to a whole family of compactification manifolds, from a dual perspective they describe a consistent subsector which is universal to a whole class of conformal field theories.

In this paper we build on the line of research outlined above and establish a universal consistent truncation of 11D supergravity based on tri-Sasakian structures in seven dimensions. We show that the 4D theory is $\mathcal N=4$ gauged supergravity with three vector multiplets. This contains as an $\mathcal N=2$ subtruncation the universal Sasaki--Einstein truncation of \cite{vargaun2}.
Since no $\mathcal N=8$ enhancement --- which would have to contain massive modes and is thus not the massless $\mathcal{N}=8$ truncation of \cite{deWitS7} --- is known, this $\mathcal N=4$ theory seems to constitute the most supersymmetric setup in which that $\mathcal{N}=2$ model can be embedded.
We also identify a different $\mathcal{N}=2$ subtruncation, corresponding to the type IIA truncations on twistor manifolds of \cite{ExploitingN=2} (see also \cite{tomasiellocosets,cosets}) in the limit of zero Romans mass.
It would be interesting to see whether the $\mathcal{N}=4$ model can be deformed so as to contain Romans mass after the subtruncation or whether this is only possible in the $\mathcal{N}=2$ formalism, although this question is outside the scope of the paper.

The possibility of a consistent truncation of 11D supergravity on 7D tri-Sasakian manifolds preserving $\mathcal N=4$ supersymmetry was suggested in \cite{vargaun3}.
Tri-Sasakian geometry (see \cite{sasakireview} for a review) admits three Killing spinors and therefore is well known to support AdS$_4$ Freund--Rubin solutions of 11D supergravity with $\mathcal N=3$ supersymmetry \cite{Acharya:1998db}. The reason why $\mathcal N=4$ supersymmetry, instead of just $\mathcal N=3$, is preserved at the off-shell level relies on the fact that tri-Sasakian structures fall in the class of SU(2)-structures. As such, they admit {\it four} globally-defined, SU(2)-invariant spinors: employing all of them, one can define an $\mathcal N=4$ truncation ansatz. The $\mathcal N=3$ background makes its appearance in our model as a spontaneously broken phase. We have therefore constructed an explicit, minimal model realizing the partial $\mathcal N=4 \to \mathcal N=3$ supersymmetry breaking phenomenon described in \cite{ShadowMult}.

The $\mathcal N=4$ ansatz turns out to be consistent due to the particularly simple SU(2)-torsion of tri-Sasakian structures, which is SU(2)-invariant and constant. Concretely, we will prove consistency of the reduction by showing that it is equivalent in form to a left-invariant truncation on a homogeneous representative of the class of tri-Sasakian manifolds, namely the coset space Sp(2)/Sp(1). We will also discuss how, in fact, all known universal consistent truncations with massive modes can be seen as left-invariant reductions on coset spaces.

We demonstrate that the reduced theory matches the general $\mathcal N=4$ gauged supergravity formalism described in \cite{SchonWeidner} (building on the ideas of \cite{samtlebenmag}).
The on-shell global symmetry group of 4D $\mathcal N=4$ supergravity coupled to $n$ vector multiplets is $G={\rm SL}(2,\mathbb{R})\times{\rm
SO}(6,n)$. The off-shell Lagrangian of \cite{SchonWeidner} contains both propagating fields, which have kinetic terms, and auxiliary fields. In particular, a specific choice
of symplectic frame for the vectors --- i.e.\ a choice of which are the propagating or electric vector potentials $\mathcal{A}_\mu^{M+}$ and which are the auxiliary magnetic potentials $\mathcal{A}_\mu^{M-}$ --- with good transformation properties under SO$(6,n)$ is made. Furthermore, there are auxiliary two-forms, while all scalars are propagating.
After some field transformations, our reduced 4D theory takes exactly the form of the $\mathcal{N}=4$ Lagrangian of \cite{SchonWeidner} with $n=3$ vector
multiplets. These manipulations, which represent a non-trivial technical point of our analysis, are required because the straightforward dimensional reduction of the action does not directly yield the electric-magnetic frame of \cite{SchonWeidner}. In order to identify the fields in our truncation ansatz with the $\mathcal N=4$ supergravity ones, we study their transformation behavior under the global symmetry group $G$ and the embedding of $G$ into the global symmetry $\text{E}_{7(7)}$ corresponding to maximal supersymmetry. For this we make use of the formalism of Exceptional Generalized Geometry \cite{walexcep} (see also \cite{hullexcep} for earlier work). Furthermore, in \cite{SchonWeidner} the gauging is described by the {\em embedding tensor}, namely a set of $G$-tensors $(\xi_{\pm M},f_{\pm MNP})$, with $f_{\pm MNP}=f_{\pm [MNP]}$, specifying how the gauge group is embedded in the global symmetry group (see \cite{samtlebenreview} for a review). We will determine
the embedding tensor and show that the gauge algebra is given by a semi-direct product of the on-shell $R$-symmetry so(3)$_R$ and a six-dimensional nilpotent algebra.

We also show how one passes from the electric-magnetic frame of \cite{SchonWeidner} to a purely electric frame, where no auxiliary vector nor two-form fields appear in the action. Because only physical degrees of freedom appear in the electric frame, it is particularly suitable for discussing questions like the spectrum of a given background and the gauge group of the model.

We indeed present some prominent AdS$_4$ solutions and the spectrum of 4D field fluctuations around them. For the supersymmetric ($\mathcal N=3$ and $\mathcal N=1$) ones, we discuss how the fields organize in supermultiplets.
Moreover, the spectrum around the non-supersymmetric Pope--Warner solution \cite{popewarner} demonstrates its instability for all tri-Sasakian manifolds, extending the result of \cite{PWunstable} beyond the $S^7$ case.

Our truncation can also be used to study some aspects of the AdS$_4$/CFT$_3$ correspondence~\cite{AdS4CTF3} with $\mathcal N=3$ supersymmetry \cite{JafferisTomasiello,Herzog:2010hf,RingsN=3}, like holographic renormalization group flows between different conformal points. In particular, we present a simple $\mathcal N=1$ subtruncation containing two AdS vacua (corresponding to a round and a squashed 7D geometry, respectively), which should be useful for this purpose. Moreover, the presence of the SO(3) factor in the gauge group makes the tri-Sasakian truncation an ideal setup for holographic applications to 3D condensed matter systems with vector order parameter. In particular, it opens the possibility to embed the holographic \hbox{$p$-wave} superconductor of \cite{Gubser:2008zu,GubserPufu}, as well as the solutions of \cite{2ndSoundSU2}, into M-theory.
Interesting in this respect is a subtruncation to minimal $\mathcal N=3$ gauged supergravity, which we also present, whose bosonic sector coincides with the Einstein--Yang--Mills model of \cite{Gubser:2008zu,GubserPufu, 2ndSoundSU2}.

We will also discuss an extended consistent reduction, based on the homogeneous seven-manifold $N^{010}$. Compared to the reduction on a generic tri-Sasakian manifold, it has an extra non-universal vector multiplet associated to the non-trivial cohomology, i.e. a Betti multiplet.

The paper is organized as follows. We start in section \ref{sec:3Sprelim} by introducing basic notions about tri-Sasakian geometry. The dimensional reduction is worked out in section~\ref{sec:DimRed}, and  matched with $\mathcal N=4$ gauged supergravity in section~\ref{sec:N4}. The AdS$_4$ solutions and their spectrum are presented in section~\ref{sec:Solutions}. Section~\ref{sec:CosetRealiz} discusses the coset realizations of our truncation as well as of the other known universal truncations. It also details the extension on $N^{010}$. Section \ref{sec:subtruncations} presents the new subtruncations we found.
Many technical details are relegated to appendices.

\section{Tri-Sasakian geometry}
\label{sec:3Sprelim}

In this section we introduce the basic notions about tri-Sasakian
manifolds that will be needed in the rest of the paper. In particular, we introduce
the tri-Sasakian structure, which consists of a set of differential forms existing on every tri-Sasakian
manifold, which we call {\em universal} forms. These are the ones used as the building blocks for our reduction ansatz.
For a thorough review on tri-Sasakian geometry we refer to \cite{sasakireview}.

Let us start with the definition of a Sasakian manifold.
A Riemannian manifold $(M,g)$ of dimension $2n+1$ is {\em Sasakian} if
its cone 
has reduced holonomy U($n+1$).

If the holonomy of the cone further reduces to SU($n+1$)
--- or in other words if the cone is a Calabi--Yau manifold --- one can show that the metric on $M$ is Einstein,
meaning that the Ricci tensor is proportional to the metric: $R_{mn} = 2n \, g_{mn} \,$.\footnote{The overall scale is conventionally
chosen such that the proportionality constant is the same as for the unit $S^{2n+1}$, the simplest example of a Sasaki-Einstein manifold.}
$M$ is then called a {\em Sasaki--Einstein} manifold. Just like Calabi--Yau
manifolds play an important role as the internal manifold for supersymmetric compactifications of 10D supergravity on Minkowski space,
Sasaki--Einstein manifolds serve as the internal manifold of supersymmetric Freund--Rubin compactifications of 11D supergravity to AdS$_4$
(for a review see \cite{KKreview}). Indeed, while a Calabi--Yau manifold supports covariantly constant spinors, Sasaki--Einstein manifolds carry Killing spinors, which satisfy
$\nabla_{m} \psi = \pm \frac{1}{2} \gamma_m \psi$,
and which play the role of supersymmetry generators. Since a Sasaki--Einstein manifold
supports two such Killing spinors, the corresponding Freund--Rubin AdS$_4$-solution has $\mathcal N=2$ supersymmetry in 4D. The Sasaki-Einstein manifold
itself has structure group SU($n$).

An alternative and equivalent characterization of a Sasaki-Einstein manifold is that
it carries a Sasaki-Einstein structure $(\xi,\eta,J,\Omega)$. Here $\xi$ is a unit Killing vector (of the metric $g$ implied
by $J$ and $\Omega$) and $\eta$ the dual one-form given by $\eta(Y)=g(\xi,Y)$. $J$ is a real two-form and $\Omega$ a complex decomposable $n$-form.
They must satisfy:
\subeq{\al{
& \d \eta  = 2 J \, , \\
& \d \Omega  = (n+1) i \,\eta \wedge \Omega \, , \\
& J \wedge \Omega  = 0 \, , \quad \iota_{\xi} J = \iota_{\xi} \Omega = 0 \, .
}}

We are now ready to define a tri-Sasakian manifold. A {\em tri-Sasakian manifold} is a Riemannian manifold such that
its cone is hyper-K\"ahler. Since all hyper-K\"ahler manifolds are Calabi--Yau, a tri-Sasakian manifold is also automatically
Sasaki--Einstein (while the converse is not true). Tri-Sasakian manifolds only exist in dimensions $4 m +3$ with $m \ge 1$.
For our purposes, the following alternative characterization is more directly useful since
it introduces the building blocks we need for our reduction ansatz.
A tri-Sasakian manifold has three unit orthogonal Killing vectors $\xi_I$, such that each of them makes
the manifold into a Sasakian manifold, and which generate the so(3)-algebra:
\eq{
\label{commxi}
[\xi_J, \xi_K] = 2 \, \epsilon^I{}_{JK} \xi_I \, ,\qquad\qquad I,J,K =1,2,3\,.
}

If we introduce the dual one-forms $\eta^I$, i.e.\ $\iota_{\xi_I} \eta^J = \delta^I{}_J$,
one can show that they satisfy
\eq{
\label{dereta}
\d \eta^I = 2 J^I - \epsilon^I{}_{JK} \eta^J \wedge \eta^K \, ,
}
with $\iota_{\xi_I} J^J=0$. The vectors $\xi_I$ define a three-dimensional foliation, called the tri-Sasakian
foliation. It can be shown that the generic leaves are either SU(2) or SO(3), and that the space of leaves $B_{\text{QK}}$
is a quaternionic K\"ahler manifold (in the regular case) or orbifold (in the quasi-regular case) with the three almost complex structures being $g^{-1} J^I$. 
The unit tri-Sasakian metric is given by
\eq{
\label{3Smetric}
\d s^2_{\text{3S}} = \d s^2(B_{\text{QK}}) + (\eta^1)^2 + (\eta^2)^2 + (\eta^3)^2 \, .
}
From eq.~\eqref{dereta} it follows
\eq{\label{derJ}
\d J^I = 2 \, \epsilon^I{}_{JK} J^J \wedge \eta^K \, .
}
In 7D ($m=1$), the case of interest for us, one can furthermore show that
\eq{\label{wedgeJ}
J^I \wedge J^J = 2 \, \delta^{IJ} \text{vol}_{\text{QK}} \, ,
}
where we introduced the standard volume $\text{vol}_{\text{QK}}$ of the 4D $B_{\text{QK}}$.
For the 7D volume form we choose the following standard orientation:
\eq{
\label{3Sori}
\frac{1}{2} J^I \wedge J^I \wedge \eta^1 \wedge \eta^2 \wedge \eta^3 = \text{vol}_{\text{QK}} \wedge \eta^1 \wedge \eta^2 \wedge \eta^3 = {\rm vol}_{\text{3S}} \, \quad \text{(no sum over $I$)}.
}
The forms $\eta^I$ and $J^I$ will be the basic ingredients to construct our truncation ansatz in the next section.

It is possible to show that a 7D tri-Sasakian manifold admits three Killing spinors, which means that one of the four
covariantly constant spinors of its 8D hyper-K\"ahler cone gets lost in the reduction along the radial direction. In this paper we will argue that the 4D theory we obtain
from reducing on a tri-Sasakian manifold has actually $\mathcal N=4$ supersymmetry,
which is spontaneously broken to $\mathcal N=3$ in the most supersymmetric AdS$_4$ phase.

By definition the vectors $\xi_I$, generating an so(3) algebra, are Killing vectors of the tri-Sasakian metric \eqref{3Smetric}.
This so(3) acts on the supersymmetry generators of the $\mathcal{N}=3$ AdS$_4$ phase so we will call it so(3)$_R$. But this is not the full space of Killing
vectors of the 7D tri-Sasakian metric, which instead spans $\text{so}(4)=\text{so}(3)_R \times \text{so(3)}_S$.
The second factor, $\text{so}(3)_S$, acts on the four-dimensional part $B_{\text{QK}}$ and is the remaining part of the so(4) that leaves the metric on $B_{\text{QK}}$
invariant. In fact, it leaves all the $J^I$ invariant. Because it leaves not only the metric, but the complete tri-Sasakian structure invariant, the SU(2)$_S$ associated to this so(3)$_S$ is identified with the structure group of the tri-Sasakian manifold. Hence the label $S$.

Interestingly, on a tri-Sasakian manifold there is an $S^2$ worth of Sasaki-Einstein structures, which we will later on use to define Sasaki--Einstein subtruncations of the type of \cite{vargaun2}.
Indeed, for every unit linear combination
of the $\xi_I$, $\xi=\alpha^I \xi_I$ with $\alpha^I \alpha^J \delta_{IJ}=1$, there is a corresponding Sasaki--Einstein structure $(\xi,\eta,J,\Omega)$, given by
\eq{\spl{
\label{subsasaki}
\eta & \,=\, \alpha_I\,\eta^I \, , \\
J & \,=\, \alpha_I\,(J^I- \tfrac 12 \epsilon^I{}_{JK} \eta^J \wedge \eta^K) \, , \\
\Omega & \,=\, (\alpha'_J \eta^J  - i \alpha''_J \eta^J) \wedge (\alpha'_K J^K - i \alpha''_K J^K ) \, ,
}}
with $\alpha_I = \delta_{IJ} \alpha^J$. To properly define $\Omega$ above, we had to choose a whole orthogonal frame $(\xi,\:\xi'=\alpha'{}^I \xi_I,\:\xi''=\xi \times \xi'=\alpha''{}^I \xi_I)$.
However, changing the choice of vector $\xi'$ orthogonal to $\xi$ merely leads to a change of phase of $\Omega$, which is usually not
considered as leading to a different Sasaki--Einstein structure. It corresponds to the remaining U(1) gauge symmetry of the Sasaki-Einstein reduction. With these definitions the positive orientation, eq.~\eqref{3Sori}, is in terms of the Sasakian structure given by $\frac{1}{6} J\wedge J\wedge J \wedge \eta<0$.

\section{The dimensional reduction}\label{sec:DimRed}

In this section we describe the reduction of 11D supergravity on a generic 7D tri-Sasakian manifold.
We introduce our reduction ansatz, which we will show to be consistent in section \ref{sec:Sp2Sp1}. After considering the
different sectors the final result for the 4D action is presented in section \ref{sec:resultaction}. In section \ref{sec:N4} we
show that it is exactly equivalent to the action of $\mathcal N=4$ gauged supergravity as described in \cite{SchonWeidner}.

The bosonic sector of 11D supergravity consists of the metric and a three-form potential $C_3$, with field-strength $G_4 = \d C_3$. The action is given by
\eq{
\label{11Dsugra}
S \,=\, \frac{1}{2\kappa_{11}^2}\int \Big\{\Big[R - \frac{1}{2}(G_4)^2\Big]\!*_{11}\!1 \,-\,\frac{1}{6} \, C_3 \wedge G_4 \wedge G_4 \Big\}\, ,
}
which is invariant under gauge transformations of the form $\delta C_3 = \d \Lambda_2$. 

\subsection{Reduction of the geometric sector}\label{sec:RedGeomSector}

Our reduction ansatz for the metric reads
\eq{\label{metricansatz}\spl{
& \d s_{11}^2 \,=\, e^{2 \varphi} \d s_4^2
 + e^{2U}  \d s^2(B_{\text{QK}}) + g_{IJ} (\eta + A_1)^I (\eta + A_1)^J \, .
}}
It contains seven 4D scalar fields $U(x),\, g_{IJ}(x)$ and three 4D vector fields $A_1^I(x)$, where $x^\mu$ are the 4D coordinates. Here $U$ controls the size of the space of leaves $B_{\text{QK}}$, while $g_{IJ}$ is the metric on the leaves. Sometimes it will be convenient
to separate the eigenvalues from the other degrees of freedom in $g_{IJ}(x)$ as follows
\eq{
\label{metriceta}
g = Q^T Q \, , \qquad\; Q = V O\,,\qquad\; V = {\rm diag}\left(e^{V_1},\,e^{V_2},\,e^{V_3} \right)  \, ,
}
where both $O(x) \in \text{SO(3)}$ and $V(x)$ contain three 4D scalar fields.
The warp factor $\varphi(x)$ in (\ref{metricansatz}) is merely a 4D Weyl rescaling factor, fixed to
\eq{
\varphi\, =\,   - \frac{1}{2}\left(4 U + V_1 + V_2 + V_3 \right)\,,
}
in order to ensure that the 4D action is in the Einstein frame.

A long calculation yields the following expression for the 11D Ricci scalar\begin{multline}\label{Ricciscalar}
R \,=\,  e^{-2 \varphi} R_{(4)} + R_{(7)} - \frac{1}{2} e^{-4 \varphi} g_{IJ} F^I_{2} \cdot F^J_{2} - 4\,(\d U)^2 \\ -\frac{1}{4} e^{-2 \varphi} D g_{IJ} \cdot D g_{KL}\, g^{IK} g^{JL}
 -2 \, e^{-2\varphi} \nabla^{\mu} \partial_\mu \varphi - 2 \, e^{-2\varphi} (\d \varphi)^2 \, ,
\end{multline}
where the $\cdot$ and $(\,)^2$ (contracting here 4D indices) are defined in \eqref{formprod}, and $R_{(4)}$ is the Ricci scalar
associated to the 4D metric $g_{(4)}$ in $\d s_4^2$. The SO(3)-covariant derivative $D$ of the metric on the leaves is
\eq{\spl{
\label{metriccov} & D g_{IJ} =\, \d g_{IJ} + 4 \, \epsilon^K{}_{L(I} A^L_{1} \, g_{J)K} \\
& \Longleftrightarrow  \, D V_I =\, \d V_I \, , \quad D O^I{}_J =\, \d O^I{}_J + 2 \, O^I{}_K \epsilon^K{}_{LJ} A^L_{1} \, ,
}}
which implies in particular that the eigenvalues are not charged, while the vector field-strength is
\eq{
\label{fieldstr} F^I_2 =\, \d A_1^I - \epsilon^I{}_{JK} A_1^J \wedge A_1^K \, ,}
and the 7D Ricci scalar, to be precise the Ricci scalar one would obtain if all the scalar fields were constant, reads
\begin{multline}
\label{7Dricci}
R_{(7)} = -2\, e^{ -2 V_1- 2 V_2 - 2 V_3} (e^{4 V_1}+e^{4 V_2}+e^{4 V_3}) +48 \, e^{-2U} \\
 +4  \, (e^{-2 V_1}+e^{-2 V_2}+e^{-2 V_3})
 -4 \, e^{-4U} (e^{2 V_1}+e^{2 V_2}+e^{2 V_3}) \, .
\end{multline}

We observe that the vectors $\xi_I$, which are Killing vectors of the Sasaki--Einstein metric \eqref{3Smetric}, are not Killing vectors of the ansatz \eqref{metricansatz}. As is standard in Kaluza--Klein reductions, they rather generate gauge transformations in 4D.
Indeed, the infinitesimal diffeomorphism generated by the $x$-dependent vector $\lambda = \lambda^I(x) \xi_I$ modifies the 11D metric as
\begin{multline}
\delta_\lambda g_{(11)} \equiv\mathcal{L}_{\lambda} g_{(11)} = 2 \, g_{IJ} \left[\mathcal{L}_{\lambda} (\eta+A_1)^I\right] (\eta+A_1)^J \\
=2 \, g_{IJ} \left[-2 \, \epsilon^I{}_{KL} \lambda^K \eta^L + \d \lambda^I \right] (\eta+A_1)^J \, .
\end{multline}
This can be interpreted as the following transformations of the 4D fields $A_1^I$ and $g_{IJ}$:
\eq{\spl{
\delta_\lambda A_1^I \,&=\,  \d \lambda^I -2 \, \epsilon^I{}_{LK} A_1^L \lambda^K \, ,\\
\delta_\lambda g_{IJ} \,&=\, 4 \, g_{L(I} \epsilon^{L}{}_{J)K} \lambda^K\,.
}}
So we conclude that the so(3)$_R$ generated by the $\xi_I$ acts as rotations
on the metric $g_{IJ}$ on the leaves, and that it is gauged by the $A_1^I$.

It is straightforward to check that the derivative in eq.~\eqref{metriccov} and the field-strength in eq.~\eqref{fieldstr} have the proper covariant structure under these SO(3)$_R$ gauge transformations. Also note that the three degrees
of freedom in $O(x)$ could in principle be gauged away, while on the other hand the eigenvalues in $V(x)$ do not transform. It follows that the 4D lagrangian \eqref{Ricciscalar} obtained from the reduction of the 11D Ricci scalar is gauge invariant,
which is of course a consequence of 11D diffeomorphism invariance.
The same will be true for the expressions that we are going to obtain by reducing the 11D form sector.

\medskip

The reduction of the Einstein--Hilbert term in the 11D supergravity action, eq.~\eqref{11Dsugra}, is now straightforward.
Plugging in the expression for the Ricci scalar \eqref{Ricciscalar} we find
\eq{
S_{\text{geom}} =\, \frac{1}{2\kappa_4^2} \int (R_4 - 2 V_{\text{geom}}) \!*\! 1 \,+\, S_{\text{geom,kin}} \, ,
}
where the 4D Planck mass is given by $\kappa_4^2 =(\text{Vol}_{M,\text{3S}})^{-1} \kappa_{11}^2$. So we see that the dependence
on the internal manifold factorizes, yielding upon integration just a factor of the volume of the manifold measured with the unit tri-Sasakian metric \eqref{3Smetric}, which contributes to the 4D Planck mass. The geometric part of the potential is given by
\begin{multline}
V_{\text{geom}}  = e^{-4U -3 V_1- 3 V_2 - 3 V_3} (e^{4 V_1}+e^{4 V_2}+e^{4 V_3}) -2  \, e^{-4 U - V_1-V_2-V_3} (e^{-2 V_1}+e^{-2 V_2}+e^{-2 V_3}) \\
 +2 \, e^{-8U - V_1-V_2-V_3} (e^{2 V_1}+e^{2 V_2}+e^{2 V_3}) -24 \, e^{-6U -V_1 -V_2 -V_3} \, ,
\end{multline}
and the kinetic terms for the scalars and vectors are
\begin{multline}\label{geomkin}
S_{\text{geom,kin}}  = \frac{1}{2\kappa_4^2} \!\int \Big[ - 12 (\d U)^2 - 4 \, \d U \cdot \d(V_1+V_2+V_3)
- \frac{1}{2} \, \big(\d (V_1 + V_2 + V_3) \big)^2 \\
 -\frac{1}{4} D g_{IJ} \cdot D g_{KL}\, g^{IK} g^{JL}
 - \, \frac{1}{2} \, e^{4U + V_1+ V_2 +V_3} g_{IJ} F_2^I \cdot F_2^J  \Big] \!*\!1\, .
\end{multline}

\subsection{Reduction of the form sector}

We can now proceed to reduce the remaining part of the 11D supergravity action, involving the form potential $C_3$. Our truncation ansatz is given by the most general expansion in the universal forms $(\eta^I,J^I)$ of the tri-Sasakian structure,
\eq{\label{C3pot}\spl{
C_3 \,=\,\, & c_3 + c_{2I} \wedge (\eta + A_1)^I + c_{1I} \wedge J^I + \frac 12 \,\tilde{c}_{1}^{\,I} \wedge \epsilon_{IJK} (\eta + A_1)^J \wedge (\eta + A_1)^K \\
& + c_{\,IJ} (\eta + A_1)^I \wedge J^J
 +  \frac 16\, \chi  \, \epsilon_{IJK}  (\eta + A_1)^I \wedge (\eta + A_1)^J \wedge (\eta + A_1)^K \, .
}}
It contains ten 4D scalar fields $(c_{IJ}(x),\chi(x))$, six vector fields $(c_{1I}(x),\tilde{c}_{1}^{\,I}(x))$, three two-form fields $c_{2I}(x)$ and a single three-form field $c_3(x)$.
Just like in the ansatz for the metric, we used the linear combinations $(\eta+A_1)$ to ensure that the 4D fields transform under SO(3)$_R$ as expected from their index structure.
Here and in the following, if we raise and lower the $I,J,\ldots$ indices we use the Kronecker-delta $\delta_{IJ}$. If instead
the metric $g_{IJ}$ should be used it will be always explicitly displayed.

We can then calculate the field-strengths $G_4 = \d C_3$. In principle, we could add an internal closed but non-exact part $G_4^{\text{flux}}$, but
this is zero for a {\em generic} tri-Sasakian manifold, since no closed, non-exact four-forms can be constructed using only universal forms. Using eqs.~(\ref{dereta}) and (\ref{derJ}) we find
\eq{\label{G4flux}\spl{
G_4 \,= &\,\, H_4 + H_{3I} \wedge (\eta + A_1)^I + \frac 12\,\tilde{H}_{2}^{\,I} \wedge \epsilon_{IJK} (\eta + A_1)^J \wedge (\eta + A_1)^K \\
+&\, H_{2I} \wedge J^I + H_{1IJ} \wedge (\eta + A_1)^I \wedge J^J
+ 4 \, ({\rm tr}\, c)\, \text{vol}_{\text{QK}} \\
+&\, \frac 16\,\d \chi \wedge \epsilon_{IJK} (\eta + A_1)^I \wedge (\eta + A_1)^J \wedge (\eta + A_1)^K \\
+&\, \left[(\chi  + \text{tr}\, c)\, \delta_{LK} - 2 \, c_{(LK)} \right] \epsilon^L{}_{IJ} (\eta + A_1)^I \wedge (\eta + A_1)^J \wedge J^K,
}}
with $\,\text{tr}\, c \equiv c_I{}^I$ and
\eq{\spl{
&  H_4  \,=\, \d c_3 + c_{2I} \wedge F_2^I \, , \\
& H_{3I} \,=\, D c_{2I} - \epsilon^K{}_{JI} \tilde{c}_{1K} \wedge F_2^J \, , \\
& \tilde{H}_{2I} \,=\, D \tilde{c}_{1I} - 2\, c_{2I} + \chi \, F_{2I} \, , \\
& H_{2I} \,=\, D c_{1I} + 2 \, c_{2I} +c_{JI} F_2^J \, , \\
& H_{1IJ} \,=\, D c_{IJ} + 2 \, (c_{1K}+\tilde{c}_{1K}) \epsilon^K{}_{IJ} \, ,}\label{Hfield-strengths}}
where the action of the SO(3)$_R$-covariant derivative $D$ on the form fields is given by
\eq{
\label{ccov}
D c_{n \, I_{1}\ldots I_{n}}\, =\, \d c_{n\,I_{1}\ldots I_{n}} + \sum_{l=1}^n 2 \, \epsilon^K{}_{JI_l} A_1^J \wedge c_{n\,I_1 \ldots K \ldots I_n}   \, .
}
The SO(3)$_R$ gauge symmetry is not the only one acting on the $c$-fields: indeed the gauge shifts of $C_3$ also induce some transformations on them. As a consequence, the SO(3)$_R$-covariant derivative $D$ will need to be extended into a covariant derivative of the full gauge group. We postpone the discussion of the full gauge symmetry of the 4D model to section \ref{sec:gauging}.

\medskip

Having defined our truncation ansatz, we can start the reduction of the form sector. The most straightforward approach is to follow the same strategy as for the geometric sector, namely
plugging the reduction ansatz for the forms into the higher-dimensional action. Again the dependence on the internal manifold
factorizes out and we end up with a 4D action.

However, this is not the most convenient approach for making explicit that the 4D theory is an $\mathcal{N}=4$ gauged supergravity. We want to prove this by showing that the action can be put in the general form presented in \cite{SchonWeidner}. The problem is that the straightforwardly reduced action cannot be directly
compared to \cite{SchonWeidner}, basically because it turns out not to be in the same electric-magnetic frame.

This is most apparent
from the observation that the straightforwardly reduced action would contain propagating 4D two-forms, i.e.\ two-forms with kinetic terms,
namely the $c_{2I}$ in \eqref{C3pot}, while in \cite{SchonWeidner} the only propagating matter fields are scalars and vectors, with the two-forms being just auxiliary fields.\footnote{A brief review of $\mathcal N=4$ gauged supergravity as described in \cite{SchonWeidner} is given in section~\ref{sec:SW} below.}
Of course, one can try to dualize these two-forms into scalar fields at the level of the action. However, while this would be easy in an ungauged theory, in the present gauged setup it is more complicated since the two-forms are massive, meaning that they are St\"uckelberg-coupled to some combinations of the vector fields. This implies that in order to perform the dualization of the two-forms one has to dualize the St\"uckelberg vectors as well. The dual vectors turn out to be massive, with the scalars dual to the two-forms playing now the role of St\"uckelberg fields \cite{louismicu1} (see also \cite{IIAonSU2} for an example in the $\mathcal N=4$ context). Now the issue is that, once these dualizations are done, one ends up with an $\mathcal N=4$ supergravity model in a purely electric frame which, though perfectly consistent and interesting in its own right, hides some of the off-shell symmetries of the theory. Because of this, a comparison with \cite{SchonWeidner} is not direct.\footnote{One would fall in the situation described in section 2.4.2 of \cite{SchonWeidner}; see section \ref{sec:gauging} below for more details.}
In particular, the formulation of \cite{SchonWeidner} includes the two-forms as non-propagating, auxiliary fields, so in order to make contact with it we should not completely dualize them away.

In order to introduce the proper dual fields and to move to the electric-magnetic frame that allows a direct comparison with \cite{SchonWeidner}, we found that working at the level of the equations of motion is more convenient than first reducing the action and then looking for the appropriate field transformations.
So in the procedure that we are going to present we manipulate the equations of motion to formulate them in the electric-magnetic frame of \cite{SchonWeidner}, and then construct the 4D action that reproduces them, directly in the proper frame. A bonus of this approach is the proof of consistency of the reduction in the form sector (the full consistency will be proved later, using a different approach). Moreover, the 11D origin of the dual fields emerges in a neat way.

We start by straightforwardly reducing the Bianchi identity and the equation of motion of the 11D form field, which read respectively
\subeq{\label{BianchiEomG4}\al{
&\d G_4 \,=\, 0 \, , \label{BianchiG4} \\
&\d *_{11} G_4 + \frac{1}{2} G_4 \wedge G_4 \,=\, 0 \, .\label{EoMG4}
}}
Plugging in eq.~(\ref{G4flux}), we find that the dependence on the internal coordinates drops out, which as announced proves consistency of the reduction in the form sector. In this way we obtain the 4D Bianchi identities and second-order equations of motion for the original $c$-fields, which are spelled out in appendix \ref{app:eombiflux}. It is easy to check that the 4D equations of motion are the ones following from the action that is obtained by direct dimensional reduction of the 11D supergravity action.

As we just discussed above this action does not match with the $\mathcal{N}=4$ formalism of \cite{SchonWeidner}.
Let us therefore introduce the dual fields. The 11D Bianchi identity and equation of motion (\ref{BianchiEomG4}) allow for defining a pair $(C_3,\widetilde C_6)$ of ``electric'' and ``magnetic'' potentials
\subeq{\al{
& G_4\, =\, \d C_3 \, , \label{defpotC3}\\
& *_{11} G_4 + \frac{1}{2} C_3 \wedge G_4 \,=\, \d \widetilde C_6 \, . \label{defpotC6}
}\label{elecmagn11Dpotentials}}
In M-theory, $C_3$ couples electrically to the M2-branes and magnetically to the M5-branes, while the converse holds for $\widetilde C_6$.
Having both $C_3$ and $\widetilde{C}_6$ is redundant, but this is exactly what we need in order to match the action of \cite{SchonWeidner}, which,
as mentioned, also includes redundant, non-propagating fields.
\begin{table}[tp]
\begin{center}
\begin{tabular}{|c|c|c|c|}
\hline
Type & \#  & $C_3$ & $\widetilde{C}_6$ \rule{0cm}{1.1em}\\
\hline
scalar/two-form & 9 & \cellcolor{propfield} $c_{IJ}$  \;&\; $a_{2}^{IJ}$\; \\
scalar/two-form & 1 & \cellcolor{propfield} $\chi$  \;&\; $\chi_2$\; \\
two-form/scalar & 3 & \;$c_{2I}$ \;& \cellcolor{propfield} $a^I$\; \\
vector/vector & 3 & \cellcolor{propfield} $c_{1I}$ \;&\; $\tilde{a}_{1I}$\; \\
vector/vector & 3 & \;$\tilde{c}_{1}^{\,I}$ \;& \cellcolor{propfield} $a_{1I}$\;\\
3-form/constant & 1 & \;$c_3$\; & \cellcolor{propfield} $k$\;\\
\hline
\end{tabular}
\caption{The $c$-fields coming from $C_3$, and their dual $a$-fields coming from $\widetilde{C}_6$.
The number of fields is denoted in the second column. The color denotes the choice of propagating
fields in order to make contact with the formalism of \cite{SchonWeidner}. The two-form dual to $\chi$ will not appear in our model since we do not have any gauging in the SL$(2,\mathbb R)$ sector.}
\label{fluxfield}
\end{center}
\end{table}

The original $c$-fields sit in the expansion of the potential $C_3$, eq.~\eqref{C3pot}, while the dual $a$-fields reside in the analogous expansion of $\widetilde{C}_6$:
\eq{\label{C6pot}\spl{
\widetilde C_6 &=\; \frac 12 \,a^I\epsilon_{IJK} (\eta+A)^J \wedge (\eta+ A)^K \wedge {\rm vol_{QK}} + a_{1I}\wedge (\eta +A)^I \wedge {\rm vol_{QK}}\\
& +\, \tilde a_{1I} \wedge J^I \wedge (\eta + A)^1 \wedge (\eta + A)^2 \wedge (\eta + A)^3 \\
&+\, \frac 12 a_{2}^{\,I}{}_{J}\, \epsilon_{IKL} (\eta+A)^K\wedge (\eta+A)^L\wedge J^J + \chi_2 \wedge {\rm vol_{QK}} + \cdots \, ,
}}
where we stopped at terms with two-form 4D part (since the higher forms will not appear in the duality relations we are interested
in). By combining eqs.~\eqref{defpotC3} and \eqref{defpotC6} and plugging in the expansions of both the $C_3$ and $\widetilde C_6$ potentials, we obtain the duality relations between the $a$-fields and the $c$-fields, which are first-order equations giving a relation between the respective (covariant) derivatives. These are worked out in detail in appendix \ref{app:dualityrel}.

We can now use the duality relations to choose which degrees of freedom in each ($c$-field, $a$-field) pair should be kept as propagating and which are auxiliary. It turns out that the proper choice is the one indicated in table \ref{fluxfield}. In a choice
between a scalar field and a two-form we saw already that the scalar should be the physical field in the formalism of \cite{SchonWeidner}. The choice of symplectic frame for the vectors is less obvious, the deeper reason for it coming
from the transformation properties under the global symmetry group as we will explain in section \ref{sec:veciden} and appendix \ref{app:detailsE77}. For now let us state that this is the choice that reproduces the correct kinetic and topological terms for the
vectors that reproduces the Lagrangian of \cite{SchonWeidner}.

As discussed in detail in appendix \ref{app:eomSWframe}, for the $a$-fields that are chosen as fundamental, the former equation of motion for the corresponding $c$-field becomes trivial when expressed in terms of the $a$-field, and plays the role
of the Bianchi identity, while the former Bianchi identity is the new equation of motion. After this procedure it is possible that the original $c$-field
is not completely dualized away, since it might appear without derivative, e.g.\ in covariant derivatives acting on other fields, in which case it cannot be replaced by
the duality relation. However, since it does not appear anymore with derivatives acting on it, it is a purely auxiliary field. At the end of this procedure we obtain the new equations of motion given in appendix \ref{app:eomSWframe}, and we can finally construct the action reproducing them, which is presented in the next subsection.

\subsection{The complete 4D action of the truncated theory}
\label{sec:resultaction}

The total action, both geometric and form sector, is then given by
\eq{\label{4Daction}
S = \frac{1}{2\kappa_4^2} \int (R_4 - 2 \, V) \!*\! 1 + S_{\text{kin,scal}} + S_{\text{kin,vec}} + S_{\text{top}} \, .
}
The scalar potential is
\al{
V  & =  e^{-4U -3 (V_1+ V_2 + V_3)} (e^{4 V_1}+e^{4 V_2}+e^{4 V_3}) -2  \, e^{-4 U - V_1-V_2-V_3} (e^{-2 V_1}+e^{-2 V_2}+e^{-2 V_3}) \nn\\
& -24 \, e^{-6U -V_1 -V_2 -V_3}  +2 \, e^{-8U - V_1-V_2-V_3} (e^{2 V_1}+e^{2 V_2}+e^{2 V_3})
+\, 4 \, e^{-12U -V_1 -V_2 -V_3} \left( \text{tr}\, c \right)^2 \nn\\
&+\, 2 \, e^{-8U -3 (V_1 + V_2 + V_3)}
\left[ (\chi  + \text{tr}\, c) \delta^{IK} - 2 \, c^{(IK)}\right] g_{IJ}\delta_{KL}
\left[ (\chi  + \text{tr}\, c) \delta^{JL} - 2 \, c^{(JL)} \right] \nn\\
&+\,  \,e^{- 12 U - 3(V_1 +V_2 + V_3)} \big[2\,c^{IJ}c_{(IJ)} -  ({\rm tr}\, c)^2 -2\, \chi \, ({\rm tr}\, c) + 3\,k \big]^2 \,.\label{4dpotentialFinal2}
}
Note that there is no dependence on the scalars $a^I$ nor on the antisymmetric
combinations $c_{[IJ]}$. However, these should not be really regarded as moduli, since they have St\"uckelberg couplings with some of the vectors and can therefore be gauged away in order to give them mass.
The scalar kinetic terms are
\eq{\spl{
& S_{\text{kin,scal}}  =  \frac{1}{2\kappa_4^2} \int *1 \, \Big[ - 12 (\d U)^2 - 4 \, \d U \cdot \d(V_1+V_2+V_3)
- \frac{1}{2} \, \big(\d (V_1 + V_2 + V_3) \big)^2 \\
& \quad -\frac{1}{4} D g_{IJ} \cdot D g_{KL}\, g^{IK} g^{JL}
 - e^{-4U} g^{IK} \delta^{JL} H_{1IJ}\cdot H_{1KL} - \frac 12 \, e^{-2 (V_1 + V_2 + V_3)} (\d\chi )^2 \\
  & - \frac{1}{2}  e^{ -2(4 U + V_1 + V_2 + V_3)}g_{IJ} \left(\mathcal Da^I + \epsilon^{IK_1K_2} c_{K_1K_3}Dc_{K_2}{}^{K_3} \right)\cdot\left(\mathcal Da^{J} + \epsilon^{JL_1L_2} c_{L_1L_3}Dc_{L_2}{}^{L_3} \right)\!\Big],\label{SkinscalGauged}
}}
while the vector kinetic terms read
\eq{\spl{
S_{\text{kin,vec}} &=  \frac{1}{2\kappa_4^2} \int \, e^{V_1+V_2+V_3}\Big[ - \frac{1}{2} \, e^{4U } g_{IJ} F_2^I \cdot F_2^J \, -\,  \delta^{IJ} H_{2I} \cdot H_{2J} \\
  -\frac 12\,  e^{-4U} &   g^{IJ}\left(\mathcal Da_{1I}  - 2\,c_{IK} \,\mathcal Dc_1^K - c_{IL}c_K{}^L F_{2}^K - \epsilon_{IKL} a^L F_{2}^K\right) \cdot \big(\mathcal Da_{1J} - \ldots\big)
\Big]\!*\!1.\label{SkinvecGauged}
}}
And finally the topological term is given by
\eq{
S_{\text{top}} =
-\frac{1}{2\kappa_4^2} \int \left[ \chi \left( \mathcal D a_{1I}\wedge F_2^I  +  \mathcal D c_{1I}\wedge \mathcal D c_{1}^I \right)  \,+\, 2\, \mathcal D \tilde c_1^I \wedge  \mathcal D \tilde a_{1I} \right].\label{StopGauged}
}
The curly covariant derivatives $\mathcal{D}$ in the above expressions are defined in \eqref{curlycovder}. In section \ref{sec:gauging} we will show that these are the proper covariant
derivatives of the full gauge group. We see that $c_{2I}$ and $\tilde c_{1}^{\,I}$ have not been completely dualized away: they appear both in the covariant derivatives and in the topological term. However they do not have kinetic terms anymore and are therefore non-propagating degrees of freedom. The same is true for the magnetic vectors $\tilde a_{1I}$ and the two-forms $a_{2}^{IJ}$.

The topological term cannot be completely reconstructed from the second-order equations of motion of appendix \ref{app:eombiflux}: its completion depends on the non-propagating
fields $\{\tilde a_1, a_2, \tilde c_1, c_2\}$ and has been deduced by imposing that their equations of motion give the corresponding duality relations. In this way the set of degrees of freedom is constrained to the physical ones. This is also in agreement with the prescription of gauged $\mathcal N=4$ supergravity. Specifically, varying the action with respect to the two-forms $a_{2}$ and $c_{2}$ gives the duality relations in \eqref{DualityFinal} between the magnetic vectors $\tilde c_{1}, \tilde a_{1}$ and the electric ones $a_{1}, c_{1}$. The $\tilde a_1$ variation yields the duality between $c_2$ and $a$, while the $\tilde c_1$ variation gives a projected duality relation between $a_2$ and the scalars.

We are now in the position of identifying our fields with the fields of $\mathcal N=4$ supergravity, which is the topic of the next section.

\section{Match with $\mathcal{N}=4$ gauged supergravity}
\label{sec:N4}

In this section we show that our 4D theory is an $\mathcal{N}=4$ gauged supergravity by putting
the action of section \ref{sec:resultaction} in the standard form of \cite{SchonWeidner}. We first
review this standard form in the next subsection before coming to the identification of the scalar
and vectors fields in the following subsections. To conclude the section we establish the embedding of
the gauge group into the global symmetry group.

\subsection{The general action for $\mathcal N=4$ gauged supergravity}
\label{sec:SW}

In this subsection we briefly review the action for $\mathcal{N}=4$ gauged supergravity
coupled to $n$ vector multiplets as described in \cite{SchonWeidner}, which makes use of the embedding tensor formalism (see \cite{samtlebenreview} for a review). In our case it will turn out that $n=3$.

The $\mathcal{N}=4$ supergravity multiplet contains as bosonic degrees of freedom a metric, six
vectors and two real scalars, while each vector multiplet contains one vector and six real scalars.

The scalar fields parameterize the coset space
\eq{
\label{gencosetman}
\frac{\text{SL}(2,\mathbb{R})}{\text{SO}(2)} \times \frac{\text{SO}(6,n)}{\text{SO}(6)\times\text{SO}(n)} \,\, .
}
The first factor can be described by a complex scalar $\tau$, with $\Im \tau >0$. To describe the
second factor we need a coset representative $L$, which is acted upon by global group elements $g\in\text{SO}(6,n)$ from the left and local elements \hbox{$h\in\text{SO}(6)\times\text{SO}(n)$}
from the right, $L \rightarrow g L h$, with the equivalence relation given by $ L \simeq L h\,,\; \forall h  $.
Alternatively the coset can be described by the symmetric positive-definite scalar matrix $M$,
defined in terms of $L$ as
\eq{
M = L L^T \, ,
}
so that using $h h^T=\bbone$ the equivalence relation is factorized out. It satisfies $M^{-1} = \eta M \eta$
where $\eta$ is the SO$(6,n)$-metric, having six negative and $n$ positive eigenvalues. To raise and lower the SO(6,$n$)-indices,
denoted $M,N,\ldots$ , we will use the metric $\eta$, while if the metric $M$ is required it will be explicitly displayed.

For the vector fields one can choose a special symplectic frame such that the subgroup SO$(1,1)\,\times\,$SO$(6,n)$
of the global symmetry group $\text{SL}(2,\mathbb{R})\times \text{SO}(6,n)$ is realized off-shell. The electric and magnetic vectors
\be
 (\mathcal{A}^{M+},\mathcal{A}^{M-}) \,\equiv\, \mathcal{A}^{M\alpha},\qquad\;\; M=1,\ldots, 6+n,\,\quad\alpha=+,-\,,
\ee
transform both in the fundamental of SO$(6,n)$ and with charge $+1$ respectively $-1$ under SO$(1,1)$. For the electric vectors $\mathcal{A}^{M+}$ there are kinetic terms in the action, while the magnetic vectors $\mathcal{A}^{M-}$ appear in the covariant derivatives as gauge fields but do not propagate.
Furthermore there are auxiliary two-forms $\mathcal{B}^{MN}$, which transform in the adjoint of the global symmetry group and are introduced to ensure gauge covariance of the vector field-strengths and closure of the gauge algebra.

The gauging is completely determined by the embedding tensor $(\xi_{\alpha M},f_{\alpha MNP})$, which specifies how the gauge group is embedded in the global symmetry group. Here $\xi_{\alpha M}$ describes the gauging in the
SL$(2,\mathbb{R})$-sector, which will be zero in our case so we will not consider it anymore.
The covariant derivatives on the scalars are then given by
\eq{
\label{covderscalars}
\mathcal D M_{MN} \,=\, \d M_{MN} + 2 \, \mathcal{A}^{P\alpha} \eta^{QR} f_{\alpha QP(M}  M_{N)R} \, ,
}
while the field-strengths for the electric and magnetic vectors are
\eq{
\label{covfieldstr}
\mathcal H^{M\pm} \,=\, \d \mathcal A^{M\pm} - \frac{1}{2} \eta^{MQ} \left( f_{\alpha QNP} \mathcal A^{N\alpha} \wedge \mathcal A^{P\pm} \mp f_{\mp QNP}\mathcal B^{NP} \right)\, .
}

The bosonic part of the $\mathcal{N}=4$ action is then given by
\eq{
S = \frac{1}{2\kappa_4^2} \int  (R - 2 V) * \!1\, +\, S_{\text{kin}}\, +\, S_{\text{top}} \, ,
}
where the kinetic terms for the scalars and the vectors are given by
\eq{\spl{
\label{SWkin}
S_{\text{kin}} = \frac{1}{2\kappa_4^2} \int & \left[\frac{1}{8} \mathcal D M_{MN} \cdot \mathcal D M^{MN} - \frac{1}{2(\Im \tau)^2} \d \tau \cdot \d \tau^* \right] *\! 1 \\
& \quad - \Im \tau \, M_{MN} \mathcal{H}^{M+} \wedge * \mathcal{H}^{N+} + \Re \tau \, \eta_{MN} \mathcal{H}^{M+} \wedge \mathcal{H}^{N+} \, .
}}
The scalar potential is
\begin{multline}
V = \frac{1}{16} \Big\{ f_{\alpha MNP} f_{\beta QRS} M^{\alpha\beta}
\left[ \,\tfrac{1}{3} M^{MQ} M^{NR} M^{PS} + \left(\tfrac{2}{3}\, \eta^{MQ} - M^{MQ} \right) \eta^{NR} \eta^{PS}\, \right] \\
 - \tfrac{4}{9} f_{\alpha MNP} f_{\beta QRS} \,\epsilon^{\alpha\beta} M^{MNPQRS} \Big\} \, ,
\end{multline}
with
\eq{
M^{\alpha\beta} = \frac{1}{\Im \tau} \left( \begin{array}{cc} 1 &-\Re\tau \\ - \Re \tau & |\tau|^2 \end{array}\right) \, ,
\qquad M_{MNPQRS} = \epsilon_{abcdef} L_M{}^a \cdots L_S{}^f \, ,\quad a=1,\ldots,6\,,
}
where $\epsilon^{+-}=1$ and $\epsilon_{abcdef}$ are respectively the epsilon-tensors in the
SL(2,$\mathbb{R}$)-space and the SO(6)-space (i.e.\ the projection on the negative eigenvalues of the SO$(6,n)$ metric $\eta$).
Finally there is a topological term, given by
\begin{multline}
S_{\text{top}}   =  \frac{1}{2\kappa_4^2} \int f_{-MNP} \mathcal{A}^{M-} \wedge \mathcal{A}^{N+} \wedge \d \mathcal{A}^{P-}
+ \frac{1}{4} f_{\alpha MNR} f_{\beta PQS} \eta^{RS} \mathcal{A}^{M\alpha} \wedge \mathcal{A}^{N+} \wedge \mathcal{A}^{P\beta} \wedge \mathcal{A}^{Q-} \\
  - \frac{1}{4} f_{+MNR} f_{-PQS} \eta^{RS} \mathcal{B}^{MN} \wedge \mathcal{B}^{PQ}
 -  f_{-MNP} \mathcal{B}^{NP} \wedge \left(\d \mathcal{A}^{M-} - \frac{1}{2} \eta^{MS} f_{\alpha SQR} \mathcal A^{Q\alpha} \wedge \mathcal A^{R-} \right).
\end{multline}

Note that both the magnetic vectors and the two-forms drop out from the action in the ungauged limit $f_{\alpha MNP}=0$.

The equations of motion from the variation of the vectors, both electric and magnetic, and the auxiliary two-forms
become:
\subeq{\al{
\delta \mathcal{A}^+: \;\quad &  \eta_{MN} * \mathcal D\mathcal G^{N-} + \frac{1}{4} f_{+MP}{}^N M_{NQ} \mathcal D M^{QP} = 0 \, , \\
\delta \mathcal{A}^-: \;\quad &  \eta_{MN} * \mathcal D\mathcal G^{N+} - \frac{1}{4} f_{-MP}{}^N M_{NQ} \mathcal D M^{QP} = 0 \, , \label{dualscal} \\
\delta \mathcal{B}: \;\quad & \mathcal{H}^{M-} - \mathcal{G}^{M-} = 0 \label{dualvec}\, ,
}}
where we defined
\subeq{\al{
\mathcal{G}^{M+} & \,=\, \mathcal{H}^{M+} \, , \\
\mathcal{G}^{M-} & \,=\, \Im \tau \, M^{MN} \eta_{NP} * \mathcal H^{P+} - \Re \tau \, \mathcal{H}^{M+} \, . \label{dualvec2}
}}
With these definitions the equations for the electric and magnetic vectors look completely similar.
However, working them out only the first equation is a real equation of motion with second-order derivatives,
while the second equation turns into a first-order duality relation between scalars and two-forms. So we see that only the electric vectors
are propagating, while the magnetic vectors are auxiliary. Similarly, the third equation coming from the auxiliary two-forms
becomes the duality relation between electric and magnetic vectors.

We remark that it is always possible to perform an appropriate symplectic transformation, rotating electric and magnetic vectors into each other, and obtain a purely electric frame where only propagating fields appear in the action. However the new electric vectors would not transform nicely under $\text{SO}(1,1)\times \text{SO}(6,n)$ as in the chosen symplectic frame above.

\subsection{Identification of the coset manifold}
\label{sec:cosetiden}

In order to put the truncated theory into the standard $\mathcal{N}=4$ supergravity form, let us for the moment turn off
the gauging and start by the identification of the coset manifold of the scalar fields.\footnote{The ungauged limit is obtained by switching off the flux $k$ and taking the forms $\eta^I$ and $J^I$ to be closed. Such closed forms with the right properties appear on the 7D manifold $K3\times T^3$ (they are the invariant forms of its SU(2) structure), so that we can obtain ungauged $\mathcal N=4$ supergravity with three vector multiplets from a truncation of fluxless 11D supergravity on $K3\times T^3$. While we cannot claim that the reduction on $K3\times T^3$ keeping all the massless moduli is consistent, the consistency of the subsector containing our three vector multiplets is guaranteed by the fact that the set $\{\eta^I, J^I\}$ is closed under the action of wedge product, Hodge star and exterior derivative, which are the operations needed in order to consistently reduce the 11D equations of motion to 4D.
}  This coset manifold must be chosen such that the kinetic terms for the scalars correspond to the
standard coset kinetic terms in \eqref{SWkin} (in the present ungauged case to be read with ordinary derivatives instead of covariant derivatives). After having dualized the two-forms $c_{2I}$ into scalars $a^I$, we end up with 20 scalar fields:
7 in the metric sector, ($U,g_{IJ})$, and 13 in the form sector, $(c_{IJ},\chi)$ coming from $C_3$ and $a^I$ coming from $\widetilde{C}_6$. This is the right amount of fields
to describe the coset manifold \eqref{gencosetman} with $n=3$.

In order to proceed we make the connection with the analysis of \cite{walexcep} on Exceptional Generalized Geometry. Reducing M-theory to 4D on a torus $T^7$ leads to
an ungauged supergravity with $\mathcal{N}=8$ and E$_{7(7)}$ global symmetry (for a review see e.g.~\cite{samtlebenreview}). The maximal compact subgroup is then SU(8)/$\mathbb{Z}_2$.
In \cite{walexcep} the notion of an exceptional generalized metric on the internal manifold is introduced, which contains the part of the ordinary metric,
the three-form $C_3$ and the dual six-form $\widetilde C_6$ with only internal indices, or in other words the part corresponding to 4D scalar fields.
Ref.~\cite{walexcep} explains how in the case of maximal symmetry the exceptional generalized metric is described by the coset
\eq{
\frac{\text{E}_{7(7)}}{\text{SU(8)}/\mathbb{Z}_2} \, .
}
The group E$_{7(7)}$ should be thought of as acting on a standard exceptional generalized metric --- by definition the one corresponding to the unit element of E$_{7(7)}$, which
we take to be the one with all scalar fields being zero --- and generating all possible exceptional generalized metrics. Acting with an element of SU(8)$/\mathbb{Z}_2$ leaves the standard metric invariant and therefore does nothing.

So we know how the identification between the 4D scalar fields and the coset manifold works in the maximally symmetric case.
In our case, however, we do not have a torus, but a less symmetric tri-Sasakian manifold, so that
E$_{7(7)}$ will partially break to a smaller bosonic symmetry group. We will argue now that this is the symmetry group \eqref{gencosetman} of $\mathcal N=4$ supergravity coupled to $n=3$ vector multiplets:
\eq{
\text{E}_{7(7)} \rightarrow \text{SL}(2,\mathbb{R}) \times \text{SO}(6,3) \, .
}

Just like in \cite{walexcep} it will be convenient to look at subgroups of E$_{7(7)}$, which can be more easily understood.
E$_{7(7)}$ contains a subgroup SL(8,$\mathbb{R}$), which in turn contains the GL(7,$\mathbb{R}$) describing
diffeomorphisms of the 7D internal manifold. Our reduction ansatz based on the tri-Sasakian structure, eqs.~\eqref{metricansatz}, \eqref{C3pot} and \eqref{C6pot},
breaks GL$(7,\mathbb{R})$ to SO$(4)\times$GL$(3,\mathbb{R})\times \mathbb{R}_0$, corresponding respectively
to rotations in the 4D part that leave
the metric $\d s^2(B_{\text{QK}})$ invariant, diffeomorphisms in the $\eta^I$-directions, and scaling of the $B_{\text{QK}}$-part.
In fact, an SU(2) factor within SO(4), which we called SU(2)$_S$ before, leaves all $J^I$ invariant, and will cancel in numerator and denominator of the coset, so we keep only the remaining SO(3). We find then that SL(8,$\mathbb{R}$)
must break as:
\eq{\label{decompSL8}
\text{SL}(8,\mathbb{R}) \rightarrow \text{SO}(3) \times \text{SL}(4,\mathbb{R}) \times \mathbb{R}_0 \, .
}
Noting that SL$(4,\mathbb{R})/\mathbb{Z}_2 \simeq$ SO$(3,3)$, we see that together with the SO(3) we
have already constructed the diagonal blocks of the SO$(6,3)$-factor of the $\mathcal N=4$ symmetry group. The $\mathbb{R}_0$ piece on the other hand is part of SL$(2,\mathbb{R})$.

To find out where the other pieces of SO$(6,3)\times$SL(2,$\mathbb{R}$) come from let us look at the decomposition of the 133-dimensional adjoint representation of E$_{7(7)}$
in representations of SL(8,$\mathbb{R}$):
\eq{\label{133split}\spl{
\bf{133} & \rightarrow \bf{63} + \bf{70} \, ,\\
\mu & \rightarrow (\mu^a{}_b,\mu_{abcd}) \, ,
}}
where $\bf{63}$ indicates the adjoint representation of SL(8,$\mathbb{R}$) and $\bf{70}$ the
representation of a four-form. The $a,b=1,\ldots, 8$ are indices of the fundamental representation of SL(8,$\mathbb{R}$).
Under \eqref{decompSL8}, the latter representation will break as
\eq{\label{4formdecomp}
\bf{70}  \rightarrow (\bf{3},\bf{6})_0 \oplus (\bf{1},\bf{1})_+ \oplus (\bf{1},\bf{1})_- }
where the subscripts denote the weight under the action of $\mathbb{R}_0$.
One can show that the ($\bf{3}$,$\bf{6}$)-part provides the off-diagonal part of SO$(6,3)$, while the $(\bf{1},\bf{1})_+\oplus(\bf{1},\bf{1})_-$ completes the SL(2,$\mathbb{R}$).

Now let us see where our scalar fields live within this coset. First any metric can be obtained by the action of GL(7,$\mathbb{R}$)
on a standard metric, so the scalar fields describing the metric, namely $U$ and $g_{IJ}$, sit in this sector, and following the breaking
pattern of GL(7,$\mathbb{R}$) it will turn out they can be found in the GL(3,$\mathbb{R}$)$\times \mathbb{R}_0$ part, and finally in SL(2,$\mathbb{R}$) and the diagonal blocks of SO(3,3).
Second, in \cite{walexcep} it is explained how a $C_3$- and $\widetilde C_6$-field
can be generated by the (exponential of the) action of $\mu_{mnp8}$ and $\mu^m{}_8$ (defined in eq.~\eqref{133split}) respectively (see eq.~(B.22) therein):
\subeq{\label{C3C6rep}\al{
\label{C3rep} &\mu_{mnp8} \,=\, \frac{1}{2} C_{3\,mnp} \, , \\
\label{C6rep} &\mu^m{}_8 \,=\, -\frac{1}{6!}\epsilon^{mm_1\ldots m_6}\, \widetilde{C}_{6\, m_1\ldots m_6} \, .
}}
Here $m,n=1,\ldots, 7$ are space indices acted upon by GL(7,$\mathbb{R}$), and $8$ is the extra SL(8,$\mathbb{R}$)-index.
It follows that the scalar fields $c_{IJ}$ and $\chi$ sit in the four-form representation $\bf{70}$, and $a^I$
in the SL(8,$\mathbb{R}$). Following their tracks through the symmetry breaking, it will turn out that $c_{IJ}$ can be found in the $(\bf{3},\bf{6})_0$ of eq.~\eqref{4formdecomp}, $\chi$
in $(\bf{1},\bf{1})_-$ and the $a^I$ in one of the off-diagonal blocks of the SO(3,3). The details of this procedure are quite technical so we relegate them to appendix~\ref{app:detailsE77}.

As a result of the analysis in the appendix, we find that the complex scalar $\tau$ parameterizing the SL(2,$\mathbb{R}$)/SO(2) factor of the scalar manifold is given by
\eq{
\tau  \,=\, \chi + i \,e^{V_1 +V_2 +V_3} \, .
}
This leads to the kinetic terms
\eq{
-\frac{1}{2\Im (\tau)^2} \d \tau \cdot \d \tau^*  = - \frac{1}{2} \big(\d(V_1+V_2+V_3)\big)^2 -\frac 12 e^{-2 V_1 -2 V_2 - 2 V_3}(\d \chi )^2  \, ,
}
which agree with the terms \eqref{SkinscalGauged} obtained from the dimensional reduction.

Turning to the SO(6,3)-part, let us choose for the SO(6,3) metric $\eta$ the explicit representation
\eq{\label{choiceeta}
\eta = \left(\begin{array}{ccc} -\bbone_3 & \bf{0} &\bf{0} \\ \bf{0} & \bf{0} & \bbone_3 \\ \bf{0} & \bbone_3 & \bf{0} \end{array} \right) \, ,
}
where every entry represents a 3$\times$3-matrix.
We find then that the SO(6,3)-part of the exceptional generalized metric corresponding to our
reduction ansatz is obtained by acting on the standard metric with the coset representative
\eq{
\label{repunrotated}
L \,=\, \mathcal{C}\mathcal{Q} \, ,
}
where $\mathcal{Q}$ and $\mathcal{C}$ contain the scalar fields in the metric and the form sector respectively.
Explicitly, they are given by
\eq{\spl{
\mathcal{Q} \,=\, & \left(\begin{array}{ccc} \bbone_3 & \bf{0} & \bf{0} \\ \bf{0} & e^{-2U}Q^{-1} & \bf{0} \\ \bf{0} & \bf{0} & e^{2U}Q^T \end{array}\right) \, , \;\;\quad \text{with $Q$ defined in eq.~\eqref{metriceta}} \, ,\\
\mathcal{C} \,=\,  &\exp  \left(\begin{array}{ccc} \bf{0} & \sqrt 2 \,{\bf c}^T & \bf{0} \\ \bf{0} & \bf{0} & \bf{0} \\ \sqrt 2 \,\bf c& \bf{a} & \bf{0} \end{array} \right)  \,, \;\qquad \textrm{with ${\bf c}_{IJ}= c_{IJ}$ and ${\bf a}_{IJ} = - \epsilon_{IJK}a^K$}\,.
}}

We remark that the expansion of the matrix exponential of $\mathcal{C}$ already breaks off at the quadratic order since the generators
are nilpotent in the $\mathcal{C}$-sector. Furthermore $Q$ could be any matrix that satisfies $Q^T Q = g\,$.
If we choose $Q$ to be upper or lower triangular we end up with the
conventional triangular gauge for the coset representative, where it is given
as a product of exponentials of nilpotent and Cartan generators. However, we
chose $Q$ as in (\ref{metriceta}) because it allows to makes the SO(3)$_R$ gauge
symmetry more manifest.

With the above expressions the matrix $M$ is then given by
\eq{
M = L L^T = \mathcal{C} \mathcal{G} \mathcal{C}^T \, , \qquad \text{with} \quad
\mathcal{G}=\mathcal{Q}\mathcal{Q}^T = \left(\begin{array}{ccc} \bbone_3 & \bf{0} & \bf{0}  \\ \bf{0} & e^{-4U} g^{-1} & \bf{0} \\ \bf{0} & \bf{0} & e^{4U} g  \end{array}\right) \, .
}

It is a long-winded, but straightforward calculation to show that with this definition of $M$ the kinetic terms for the scalars in eq.~\eqref{SWkin} at the ungauged level agree perfectly with eq.~\eqref{SkinscalGauged}.

\subsection{Identification of electric and magnetic vectors}
\label{sec:veciden}

After identifying the coset manifold for the scalars and constructing the scalar $\tau$  and the matrix $M$, we can deduce the identification of the electric vectors from comparing
the general $\mathcal N=4$ vector kinetic terms \eqref{SWkin} to the kinetic terms in \eqref{SkinvecGauged} and the instanton terms proportional to $\chi$
in \eqref{StopGauged}. We find that they match provided we take
\eq{
\mathcal A^{I+} = c_{1}^{\,I} \,,\qquad  \mathcal  A^{(3+I)+} = -\frac{1}{\sqrt 2}\, a_{1}^{\,I}\,,\qquad \mathcal A^{(6+I)+} = \frac{1}{\sqrt 2}\,A_1^I \, .
}
Furthermore, from the duality relations between electric and magnetic vectors, eq.\ \eqref{dualvec} together with \eqref{dualvec2}, we find
the following identification of the magnetic vector fields
\eq{
\mathcal A^{I-} = -\tilde{a}_{1}^{\,I} \, , \qquad \mathcal A^{(6+I)-} = \tilde A^{I}_{1} \, , \qquad \mathcal A^{(6+I)-} = \frac{1}{\sqrt 2}\tilde c_{1}^{\,I} \, .
}
It turns out that the magnetic duals $\tilde A_{1}^I$ of the vectors in the metric sector never appear in our reduction.
It would be interesting to work out a dimensional reduction in which also these fields, which would come from a dual internal metric coupling electrically to the KK monopole and possibly arise from reducing the $h_{a_1\ldots a_8}{}^b$ tensor of \cite{WestE11}, participate in the gauging.

The matching of the kinetic terms and the instanton terms of the electric vectors would not have worked if we had not first dualized the $\tilde{c}_{1}^{\,I}$, i.e.\ if we had not treated them as magnetic vectors. There is an independent way of seeing that (in the frame of \cite{SchonWeidner}) these are magnetic, namely studying again the embedding of SL(2,$\mathbb{R}$)$\times$SO(6,3) in E$_{7(7)}$. We have already remarked that ref.\ \cite{SchonWeidner} uses
a special symplectic frame where the electric and magnetic vectors $(\mathcal{A}^{M+},\mathcal{A}^{M-})$
transform both in the fundamental of SO$(6,3)$ --- and in particular do not mix --- and with charge $+1$ respectively $-1$ under SO$(1,1)\subset$ SL(2,$\mathbb{R}$). In appendix \ref{app:detailsE77} we work out these charges for the vectors in our truncation, determining which ones are electric and which are magnetic. In particular, we find that the $\tilde c_{1}^{\,I}$ transform with a minus sign and should therefore be magnetic.

\subsection{The gauging}
\label{sec:gauging}

If we compare the covariant derivatives of the scalars \eqref{covderscalars} and the covariant field-strengths \eqref{covfieldstr} with the results
from the reduction, eqs.~\eqref{SkinscalGauged} and \eqref{SkinvecGauged}, we find exact agreement by choosing the following gaugings:
\eq{\spl{
& f_{+\,(I+3)(J+6)(K+6)} \,=\, -f_{+\,IJ(K+6)} \,=\, 2 \sqrt{2} \, \epsilon_{IJK} \, , \\
& f_{+\,(I+6)(J+6)(K+6)} \,=\, 6\sqrt{2}\, k  \, \epsilon_{IJK} \, , \\
& f_{-\,I(J+6)(K+6)} \,=\, - 4 \, \epsilon_{IJK} \, ,
}\label{EmbTensor}}
and the following identification of the auxiliary two-forms
\eq{\spl{
& \mathcal B^{(6+I)\,(6+J)} \,=\, \frac 12\left(\epsilon^{IJK} c_{2K} + A_1^{[I}\wedge \tilde c_1^{J]} \right)\, , \\
& \mathcal B^{I\,(6+J)} \,=\, - \mathcal B^{(6+J)\,I} = \frac{1}{\sqrt{2}} \left(a_2^{IJ} - \frac{1}{2} A_1^I \wedge \tilde{a}_1^J \right)\, .
}}
In particular, we find for
the covariant electric field-strengths of $\mathcal N=4$ supergravity
\bea
\label{covfieldstrel}
\mathcal H^{I+} &=& \mathcal D c_{1I} = D c_{1I} + 2 \, c_{2I} \, , \nn \\ [2mm]
-\sqrt 2\,\mathcal H^{(3+I)+} &=& \mathcal D a_{1I} = Da_{1I}+\epsilon_{IJK}\big(- 2\,c_1^J \wedge c_1^K - 2\,c_1^J \wedge \tilde c_1^K + 3k A^J\wedge A^K +4 \,a_2^{JK}\big), \nn \\ [2mm]
\sqrt 2\,\mathcal H^{(6+I)+} &=& F_2^I \equiv DA_1^I\,,
\eea
while the six relevant magnetic field-strengths are
\bea
\label{covfieldstrmag}
- \mathcal{H}^{I-} &=& \mathcal D\tilde a_{1I} = D\tilde a_{1I} + \epsilon_{IJK}\left(\tilde c_{1}^J\wedge c_1^K + \tilde c_{1}^J\wedge\tilde c_1^K + 2\,a_{2}^{JK} \right)\, , \nn \\ [2mm]
\sqrt 2\,\mathcal{H}^{(6+I)-} &=& \mathcal D\tilde c_{1I} = D \tilde c_{1I} - 2 \, c_{2I}\, .
\eea
In the above the curly derivatives $\mathcal{D}$ are the covariant derivatives associated to full gauge group, defined in eq.~\eqref{curlycovder}.
With these covariant field-strengths the duality relations, eqs.~\eqref{dualvec} and \eqref{dualvec2}, agree perfectly with the duality relations coming from
the reduction, eqs.~\eqref{DualityFinalvec1} and \eqref{DualityFinalvec2}. The duality relation between two-forms and scalars, eq.~\eqref{dualscal}, matches with \eqref{DualityFinalscal}.
Finally, the scalar potential \eqref{4dpotentialFinal2} and the topological term \eqref{StopGauged} also agree with the results from the reduction.

\medskip

It can be shown that for every consistent gauging one can perform a symplectic rotation
such that only the electric vector fields serve as gauge fields \cite{samtlebenmag}. In our setup a symplectic
rotation is a change of basis for the electric and magnetic vectors of the form
\eq{
\label{symprot}
\widetilde{\mathcal A}^{\,M,\alpha} = T^{M,\alpha}{}_{N,\beta} \mathcal{A}^{N,\beta} \, ,
}
satisfying $T^T \Omega T = \Omega$ where $\Omega$  is the symplectic form and
the matrix $T$ is obtained after lowering the indices in the magnetic sector:
\eq{
\Omega = \bbone_{9 \times 9} \otimes i \sigma_2 \, , \qquad T = \left(\begin{array}{cc} T^+{}_+\rule{1em}{0em} & T^+{}_- \eta \\ \eta T^-{}_+ & \eta T^-{}_- \eta \end{array} \right) \, .
}
Concretely, for our gauging the purely electric frame is reached by the transformation
\eq{\spl{
\widetilde{\mathcal A}^{\,I+} = \,&\, \cos\alpha\, \mathcal A^{I+} + \sin\alpha\, \mathcal A^{(6+I)-} \,=\, \frac{1}{\sqrt{3}} (c_{1}^{\,I} + \tilde{c}_{1}^{\,I}) \, ,\\
\widetilde{\mathcal A}^{\,(3+I)+} = \,&\,  \cos\alpha\, \mathcal A^{(3+I)+} - \sin\alpha\, \mathcal A^{I-} \,=\, - \frac{1}{\sqrt{6}} (a_1^I - 2 \, \tilde{a}_1^I)\, , \\
\widetilde{\mathcal A}^{\,(6+I)+} = \,&\,  \mathcal A^{(6+I)+} \,=\, \frac{1}{\sqrt{2}} A_1^I\, , \\
\widetilde{\mathcal A}^{\,I-} = \,&\, \cos\alpha\, \mathcal A^{I-} + \sin\alpha\, \mathcal A^{(3+I)+} \,=\, -\frac{1}{\sqrt{3}} (a_1^I + \tilde{a}_1^I) \, , \\
\widetilde{\mathcal A}^{\,(3+I)-} = \,&\,  \mathcal A^{(3+I)-} \,=\, \tilde A^{I}_{1} \, , \\
\widetilde{\mathcal A}^{\,(6+I)-} = \,&\,  \cos\alpha\, \mathcal A^{(6+I)-} - \sin\alpha\, \mathcal A^{I+} \,=\, \frac{1}{\sqrt{6}} (\tilde{c}_{1}^{\,I} - 2 \, c_{1}^{\,I})\, .
}}
where $\cos \alpha = 1/\sqrt 3$, $\sin \alpha = \sqrt {2 / 3}$ and $\alpha$ is a de Roo--Wagemans angle.

The gauge group generators then become
\eq{\spl{
& \widetilde{X}_{M+}  \,=\, (T^{-1})^{Q \alpha}{}_{M+} \, f_{\alpha,QNP} \, t^{NP} \,\equiv\, \widetilde{X}_M\, , \\
& \widetilde{X}_{M-} \,=\, (T^{-1})^{Q\alpha}{}_{M-} \, f_{\alpha,QNP} \, t^{NP} \,=\, 0 \, ,
}}
where $t^{NP}$ are the generators of the global symmetry group SO$(6,3)$, satisfying the algebra\footnote{In the vector representation we can take $(t^{MN})_P{}_{\rule{0em}{1em}}^Q= \eta_{\rule{0em}{1em}}^{Q[M} \delta^{N]}_P$.}
\eq{
[t^{MN}, t^{PQ}] \;=\; t^{M[P} \eta^{Q]N} - t^{N[P}\eta^{Q]M} \,.
}
We find that in the new frame the covariant derivatives
\eq{
\mathcal D \,=\, \d - \widetilde{\mathcal A}^{\,\alpha M} \widetilde{X}_{\alpha M} \,=\, \d - \widetilde{\mathcal A}^{\,+M} \widetilde{X}_M \,  ,
}
indeed do not contain magnetic vectors anymore. In the end the auxiliary two-forms as well
as the new magnetic vectors $\widetilde{\mathcal A}^{\,-M}$ completely drop out of the Lagrangian.

Let us now identify the gauge algebra, which will be conveniently done in the purely electric frame. If we introduce a new basis $X_M$ for the gauge generators given by
\eq{\spl{
& X_{I} = \frac{1}{4\sqrt{2}} \widetilde{X}_{6+I} - \frac{\sqrt{6}}{8} k \, \widetilde{X}_{3+I} \, , \\
& X_{3+I} = -\frac{1}{4\sqrt{6}} \widetilde{X}_{I}  \, , \\
& X_{6+I} = \frac{1}{4\sqrt{6}} \widetilde{X}_{3+I} \, ,
}}
we find the following non-zero commutators
\eq{\spl{
& [X_I, X_J] \,=\, \epsilon^K{}_{IJ} X_K \, , \\
& [X_{3+I},X_{3+J}] \,=\, \epsilon^K{}_{IJ} X_{6+K} \, \\
& [X_I, X_{3+J}] \,=\, \epsilon^K{}_{IJ} X_{3+K} \, , \qquad [X_I, X_{6+J}] \,=\, \epsilon^K{}_{IJ} X_{6+K} \, .
}}
The first line describes the algebra so(3), while the second line corresponds to the six-dimensional
nilpotent algebra (6,3) of \cite{magnin} (which is nilpotent algebra 3.5 of table 4 of \cite{granascan}).
The third line indicates that the nil$_{(6,3)}$-ideal transforms as $\bf{3}\oplus\bf{3}$ under so(3).
So in the end we find for the gauge algebra the semi-direct product
\eq{
\text{so(3)} \ltimes_{\bf{3}\oplus\bf{3}} \text{nil}_{(6,3)} \, .
}

Since in the purely electric symplectic frame only fundamental, propagating degrees of freedom appear, it is the appropriate setup for discussing some physical questions like the spectrum of fluctuations around a given background, which will be the subject of section \ref{sec:spectrum} below.

\section{AdS solutions and their spectra}\label{sec:Solutions}

We will show in section \ref{sec:Sp2Sp1} that the tri-Sasakian truncation we have constructed
is consistent, so that all solutions of the 4D action \eqref{4Daction} lift to solutions of 11D supergravity.
In this section we focus on vacuum solutions and leave more complicated solutions to further work.
We also calculate the spectrum of the scalars and the vectors around the supersymmetric solutions, and
show how they combine in supermultiplets. It will turn out that the spectrum provides inspiration for novel
consistent subtruncations which we will present in the next section.

\subsection{AdS$_4$ vacua}
\label{sec:vacsol}

To construct vacuum solutions, i.e.\ maximally symmetric solutions with constant scalars and vanishing vectors, we just need to
look for extrema of the scalar potential, given in eq.~\eqref{4dpotentialFinal2}. The metric then just satisfies the standard 4D Einstein equation with cosmological constant term $\Lambda = V|_{\text{sol}}$. In all the solutions below we have $\Lambda < 0$ so that the 4D geometry
is AdS$_4$. Let us first list these solutions and discuss them afterwards.

All SO(3)$_R$-invariant vacuum solutions are:
\begin{enumerate}
\subeq{
\item $\mathcal{N}=3$ supersymmetric tri-Sasakian solution ($k>0$) and skew-whiffing ($k<0$):
\label{N3susysol}
\eq{\label{3Ssol}\spl{
& e^U = e^{V_1}=e^{V_2}=e^{V_3}= |k|^{1/6}\, ,\quad c_{(IJ)}= \chi = 0\, ,  \\
& f = - 6 \, \text{sign}(k) |k|^{-1/6} \, , \quad \Lambda = - 12 |k|^{-3/2} \, .
}}
Here $k$ corresponds to the magnetic flux (see eq.~\eqref{kflux}), which is defined such that for $k=1$
we obtain the tri-Sasakian solution with standard normalization, i.e.\ the one at the origin of the scalar manifold.
\item $\mathcal{N}=1$ supersymmetric {\em squashed} solution with weak G$_2$ holonomy ($k<0$) and skew-whiffing ($k>0$):
\label{sol2}
\eq{\label{squashedSol}\spl{
& \frac{1}{\sqrt{5}} \, e^U = e^{V_1} = e^{V_2}= e^{V_3}= \left(\frac{|k|}{15}\right)^{\!1/6}\,,\quad c_{(IJ)}= \chi = 0 \, , \\
& f = - \text{sign}(k) \frac{18}{5} \left(\frac{15}{|k|} \right)^{1/6} \, , \quad \Lambda = - \frac{324}{25}\left(\frac{3}{5}\right)^{1/2}|k|^{-3/2}
}}
\item Englert solution corresponding to solution \ref{sol2}:
\eq{\spl{
& \frac{1}{\sqrt{5}} \, e^U = e^{V_1} = e^{V_2}= e^{V_3} = \left( \frac{4k}{75} \right)^{\!1/6}\,, \quad
 c_{(IJ)} =  \pm\left( \frac{k}{3}\right)^{\!1/2} \!\delta_{IJ}, \\
 & \chi =  \pm \frac{1}{5}\left( \frac{k}{3}\right)^{\!1/2}, \quad
 f = -6 \left(\frac{2}{5}\right)^{\!2/3}\! \left( \frac{3}{k}\right)^{\!1/6} \, , \quad
 \Lambda =  - \frac{27\sqrt 3}{4}k^{-3/2} \, ,
}}
which only exists for $k>0$.
}
\end{enumerate}
In the above we also displayed the Freund--Rubin parameter $f$, which is defined by
\eq{
H_4 = e^{4 \varphi} f \text{vol}_4 \, ,
}
where $H_4$ is the purely external part of $G_4$, defined in eq.~\eqref{G4flux}.

Let us now comment on the interpretation of these solutions. Since the tri-Sasakian geometry supports three Killing spinors (see e.g.~\cite{sasakireview}), the tri-Sasakian solution preserves three supersymmetries. This means that the $\mathcal{N}=4$ of the 4D theory is spontaneously broken to $\mathcal{N}=3$. We will not analyse the supersymmetry conditions to show this explicitly, but we will demonstrate that the scalars and vectors around this vacuum arrange into $\mathcal{N}=3$ multiplets.
Apart from the supersymmetric solution, there is also the so-called
skew-whiffed solution, which is obtained by changing the sign of $f$ and $k$, and which breaks all supersymmetry \cite{skewwhif}.

The second solution is based on the second Einstein metric with weak G$_2$ holonomy, which exists on
every tri-Sasakian manifold (see e.g.\ section 2.4 of \cite{sasakireview}). In the case of $S^7$, for instance, this would be
the squashed sphere geometry. Since this geometry has only one Killing spinor, the solution has $\mathcal{N}=1$. The supersymmetry that is preserved here is the one that is spontaneously broken in the $\mathcal N=4\to \mathcal N=3$ vacuum, which explains how this solution can select one supersymmetry generator and still not break the SO(3)$_R$ invariance.\footnote{In \cite{tomasiellocosets} (see also \cite{cosets}) it has been shown that (at least in the case of regular tri-Sasakian manifolds, in which case the associated 6D twistor space, where the type IIA theory lives, is a manifold) there exists a family of type IIA solutions which interpolates between the type IIA reduction of the $\mathcal{N}=3$ solution
on the one hand and the squashed $\mathcal{N}=1$ solution on the other hand.}

Finally, the third solution is the so-called
Englert solution \cite{englert} (see also section 10.2 of \cite{KKreview}) associated to the squashed $\mathcal{N}=1$ solution.
The Englert solution associated to the supersymmetry generator of a supersymmetric M-theory solution is obtained from this supersymmetric solution by turning on an internal expectation value of the field-strength $G_4$ that is a bilinear of the supersymmetry generator. The Englert solution has an opposite sign for the Freund--Rubin parameter with respect to the supersymmetric
solution (which is why we know $k<0$ corresponds to the supersymmetric solution and $k>0$ to the skew-whiffed solution). The Englert solution itself breaks all supersymmetry.

Apart from the SO(3)$_R$-invariant vacuum solutions, one can also look for solutions with less global symmetry. In the next section we will show that our tri-Sasakian reduction admits as a subtruncation the Sasaki--Einstein reduction of \cite{vargaun2} (see eq.
\eqref{sasakitrunc} below). So in particular we recover their solutions. As for vacuum solutions one finds all standard solutions on Sasaki--Einstein structures, namely the supersymmetric solution (and skew-whiffing), which is
in our case just solution 1 above, together with the Pope-Warner solution \cite{popewarner} and the Englert solution associated to one of the three Killing spinors
of the $\mathcal{N}=3$ solution. The latter two read (with the notation of \eqref{sasakitrunc}):
\subeq{\label{nonsusysol}\al{
\text{Pope-Warner:}\qquad& \sqrt{2} \, e^U = e^{V} = (- 4 k)^{1/6} \, , \quad |c_{\Omega}|= \sqrt{-k/2}  \, , \quad c_0 =0 \, , \nonumber \\
& f = 2 \sqrt{2} \left(-\frac{2}{k}\right)^{\!1/6} \! , \quad \Lambda= - 16 (-k)^{-3/2} \, , \quad k<0\,, \label{popewarner} \\
\text{Englert:}\qquad& e^U = e^{V} = \left(- \frac{4}{5} k\right)^{\!1/6} \, , \quad |c_{\Omega}|= \pm c_0 = \sqrt{-k/5}  \, , \nonumber \\
& f = 2^{5/3} \left(-\frac{5}{k}\right)^{\!1/6} \! , \quad \Lambda= - \frac{25 \sqrt{5}}{4} (-k)^{-3/2} \, , \quad k<0\, .\qquad   \label{englertN3}
}}
These solutions spontaneously break the SO(3)$_R$ completely. It follows that by acting with this group we obtain a full SO(3) worth of solutions. We also remark that since the Englert solution has positive Freund--Rubin parameter, we can indeed decide that the $\mathcal{N}=3$ supersymmetric solution in \eqref{3Ssol} is the one with negative $f$ and thus $k>0$. Note in particular that the $\mathcal{N}=3$ supersymmetric solution has a sign for $k$ opposite to the $\mathcal{N}=1$ solution \cite{squashedpagepope}.

It is difficult to find extrema of the scalar potential in complete generality. We did not find any further vacuum solutions, but we cannot claim our scan is complete as far as non SO(3)$_R$-invariant solutions are concerned.

\subsection{Spectrum}\label{sec:spectrum}

Let us now calculate the mass spectrum around these solutions. We consider fluctuations of the fields around the vacuum and
break the equations of motion off at linear order so that they take the form:
\eq{\spl{
& \Box \, \delta \Phi^A - (m^2_\Phi)^A{}_B \delta\Phi^B = 0 \, , \\
& \Box \, \delta \widetilde{\mathcal{A}}^M_\mu +\nabla_{\mu} \nabla^\rho \, \delta \widetilde{\mathcal{A}}^M_\rho - (m^2_\mathcal{A})^M{}_N \delta \widetilde{\mathcal{A}}^N_\mu = 0 \, ,
}}
where the $\delta \Phi^A$ ($A=1,\ldots,20$) are the fluctuations of the scalars, and $\delta \widetilde{\mathcal{A}}^M_\rho$ the fluctuations
of the vectors (for which we found it convenient to work in the purely electric frame). For the $\mathcal{N}=4$ action of section \ref{sec:SW} we find in general
the following expressions for the mass-squared matrices:
\subeq{\al{
 (m^2_\Phi)^A{}_C = &\,\, 2 \, G^{AB} \frac{\partial^2 V}{\partial \Phi^B \partial \Phi^C} \, , \\
 (m^2_\mathcal{A})^M{}_P = & \,\frac{1}{4\,\Im \tau} \left(T^{M+}{}_{U+} M^{UN} f_{+NTL} + |\tau|^2 T^{M+}{}_{U-} M^{UN} f_{-NTL}\right) \nonumber \\
& \left(f_{\alpha RSQ} (T^{-1})^{R\alpha}{}_{P+}\right)
 \left(M^{TS} M^{LQ} - \eta^{TS} \eta^{LQ}\right) \, ,
}}
where $G_{AB}$ is the metric on the scalar manifold, which we can find from the kinetic terms in eq.~\eqref{SWkin}, $T^{M\alpha}{}_{N\beta}$
describes the transformation that brings us to the purely electric frame for the vectors defined in eq.~\eqref{symprot} and $f_{\alpha MNP}$ is the embedding tensor
defined in eq.~\eqref{EmbTensor}.

The results for the spectrum of the scalar fluctuations around the $\mathcal{N}=3$ and $\mathcal{N}=1$ solutions are given in respectively tables \ref{spectrumN3scal} and \ref{spectrumN1scal}, while the spectrum of the vector fluctuations in both cases is the same and given in table \ref{spectrumvec}.
In each case the SO(3)$_R$ representation is indicated. We also add the scaling dimension of the operator in the dual SCFT, which for an $l$-form field is given by\footnote{When $3 m^2/|\Lambda|\le -5/4$ there are two possible normalizable modes and hence
the choice of sign. Comparing with the structure of $\mathcal{N}=3$ multiplets and later on the embedding into the $S^7$ spectrum
will fix this in all cases to the plus sign.}
\eq{
\Delta = \frac{3}{2} \pm \frac{\sqrt{(3-2l)^2 + 12\,m^2/|\Lambda|}}{2} \, .
}
Note also that we do not display the St\"uckelberg fields $c_{[IJ]}$ and $a^I$, which have been eaten by six of the vectors giving them mass.

\begin{table}[tp]
\begin{center}
\rowcolors{4}{white}{gray!50}
\begin{tabular}{|c|c|c|c|c|c|}
\hline
\multirow{2}{*}{Mass eigenstate} & \multirow{2}{*}{SO(3)$_R$} & \multicolumn{2}{c|}{susy} & \multicolumn{2}{c|}{skew}\\\cline{3-6}
 & & $\frac{3m^2}{|\Lambda|}$  & $\Delta$ & $\frac{3 m^2}{|\Lambda|}$ & $\Delta$ \rule{0cm}{1.1em}\\
\hline
$8 \, \delta U + 3 \, \text{tr}\, \delta \ln g$ & \bf{1} & 18 & 6 & 18 & 6 \\
$-3 \, \delta \chi + 2 \, \text{tr} \, \delta c$ & \bf{1} & 10 & 5 & -2 & 2 \\
$\delta c_{(IJ)} - \frac{1}{3} \delta_{IJ} \, \text{tr} \, \delta c $ & \bf{5} & 10 & 5 & -2 & 2 \\
$-6\,\delta U + \text{tr}\, \delta \ln g$ & \bf{1} & 4 & 4 & 4 & 4 \\
$\delta g_{IJ}- \frac{1}{3} \delta_{IJ} \, \text{tr} \, \delta g$ & \bf{5} & 4 & 4 & 4 & 4\\
$\delta \chi + \text{tr} \, \delta c$ & \bf{1} & 0 & 3 & 18 & 6\\
\hline
\end{tabular}
\caption{Spectrum of the scalars around the $\mathcal{N}=3$ tri-Sasakian solution ($k>0$) and its skew-whiffed counterpart ($k < 0$).}
\label{spectrumN3scal}
\end{center}
\end{table}
\begin{table}[tp]
\begin{center}
\rowcolors{4}{white}{gray!50}
\begin{tabular}{|c|c|c|c|c|c|}
\hline
\multirow{2}{*}{Mass eigenstate} & \multirow{2}{*}{SO(3)$_R$} & \multicolumn{2}{c|}{susy} & \multicolumn{2}{c|}{skew}\\\cline{3-6}
& & $\frac{3m^2}{|\Lambda|}$  & $\Delta$ & $\frac{3 m^2}{|\Lambda|}$\rule{0cm}{1.1em} & $\Delta$ \\
\hline
$8 \, \delta U + 3 \, \text{tr}\, \delta \ln g$ & \bf{1} & 18 & 6 & 18 & 6 \\
$5 \, \delta \chi + 2 \, \text{tr} \, \delta c$ & \bf{1} & 10 & 5 & -2 & 2\\
$\delta g_{IJ}- \frac{1}{3} \delta_{IJ} \, \text{tr} \, \delta g$ & \bf{5} & 52/9 & 13/3 & 52/9 &13/3 \\
$\delta c_{(IJ)} - \frac{1}{3} \delta_{IJ} \, \text{tr} \, \delta c $ & \bf{5} & 10/9 & 10/3 & 190/9 & 19/3\\
$-15 \, \delta \chi + \text{tr} \, \delta c$ & \bf{1} & -8/9 & 8/3 & 10/9 & 10/3\\
$-6\,\delta U + \text{tr}\, \delta \ln g$ & \bf{1} & -20/9 &5/3 & -20/9 & 5/3 \\
\hline
\end{tabular}
\caption{Spectrum of the scalars around the $\mathcal{N}=1$ weak G$_2$ solution ($k<0$) and its skew-whiffed counterpart ($k>0$).}
\label{spectrumN1scal}
\end{center}
\end{table}
\begin{table}[tp]
\begin{center}
\rowcolors{2}{white}{gray!50}
\begin{tabular}{|c|c|c|c|c|}
\hline
Mass eigenstate & SO(3)$_R$ rep. & $\frac{3m^2}{|\Lambda|}$\rule{0cm}{1.1em} & $\Delta$ \\
\hline
$3k \,\delta A^I - (\delta a_1^I -2 \, \delta \tilde{a}_1^I) $ & \bf{3} & 12 & 5\\
$\delta c_{1}^{\,I} + \delta \tilde{c}_{1}^{\,I}$ & \bf{3} & 6 & 4\\
$k \, \delta A^I + (\delta a_1^I -2 \, \delta \tilde{a}_1^I)$ & \bf{3} & 0 & 2\\
\hline
\end{tabular}
\caption{Spectrum of the vectors around both the $\mathcal{N}=3$ tri-Sasakian solution and the $\mathcal{N}=1$ solution, and their skew-whiffed counterparts.}
\label{spectrumvec}
\end{center}
\end{table}

In order to understand the supermultiplet structure of the spectrum, we will compare with \cite{Fre':1999xp}, where the $\mathcal{N}=3$ supermultiplets for a theory
on AdS$_4$ are classified.  We find that we can arrange the fluctuations around the $\mathcal{N}=3$ vacuum into a short massless graviton multiplet SD$(2, 3/2, 0|3)$ (table 3 of that paper)
and a long massive gravitino multiplet SD$(3/2,3,0|3)$ (table 2) associated to the broken supersymmetry (as was already predicted in \cite{vargaun2}). Here we used the notation of \cite{Fre':1999xp} for the irreducible unitary representations of the supergroup Osp($3|4$) of an $\mathcal{N}=3$ supersymmetric theory on AdS$_4$, where with
\eq{
\text{SD}(s_{\text{max}},\Delta_0,J_0|3) \, ,
}
a multiplet with maximal spin $s_{\text{max}}$, Clifford vacuum scaling dimension $\Delta_0$ and spin of the SO(3)$_R$ representation $J_0$ is denoted. The bosonic content of the massless graviton multiplet
SD$(2, 3/2, 0|3)$ and the gravitino multiplet SD$(3/2,3,0|3)$ is given respectively by:
\eq{\spl{
&\text{metric}: \, (3,{\bf{1}}) \, , \quad \text{vectors}: \,  (2,{\bf{3}}) \, , \\
&\text{vectors}: \, (5,{\bf{3}}), (4,{\bf{3}}) \, , \quad \text{scalars}: \,  (6,{\bf{1}}),(5,{\bf{1}}+{\bf{5}}),(4,{\bf{1}}+{\bf{5}}),(3,{\bf{1}}) \, ,
}}
where we indicated the scaling dimension and SO(3)$_R$ representation. Comparing with tables \ref{spectrumN3scal} and \ref{spectrumvec} we find that the massless vectors fill out
the graviton multiplet together with the metric, while the massive vectors together with the scalars fill out the gravitino multiplet.

Before coming to the spectrum around the $\mathcal{N}=1$ supersymmetric solution, we note first that the irreducible unitary representations of $\mathcal{N}=1$ AdS supersymmetry have been classified in \cite{heidenreich} and reviewed for instance in section 3.2 of \cite{KKreview}, to which we refer for the exact content of these multiplets. We find that the fluctuations arrange into the following multiplets:
\begin{itemize}
\item a massless graviton multiplet SD$(2,3|1)$, containing the metric;
\item three massless vector multiplets SD$(1,2|1)$, containing the massless vectors of table \ref{spectrumvec}, and transforming as a $\bf{3}$ under SO(3)$_R$;
\item three massive gravitino multiplets SD$(3/2,4|1)$, containing pairwise the massive vectors with $\Delta=4$ and $\Delta=5$ of table \ref{spectrumvec}, and also transforming as a $\bf{3}$ under SO(3)$_R$;
\item various Wess-Zumino multiplets SD$(0,\Delta_0|1)$ in $\bf{1}$ and $\bf{5}$ representations of SO(3)$_R$, containing pairwise the massive scalars of table \ref{spectrumN1scal}
with $\Delta=\Delta_0$ and $\Delta=\Delta_0+1$.
\end{itemize}

As for the spectrum of the non-supersymmetric solutions, we found that the Pope-Warner solution, eq.~\eqref{popewarner}, and the Englert solution associated to the $\mathcal{N}=3$ solution,
eq.~\eqref{englertN3}, are unstable since they have scalar modes violating the Breiten\-lohner-Freedman bound,
\eq{
m^2 < - \frac{3}{4} |\Lambda| \, ,
}
within the truncation.

For the Pope--Warner solution this extends the result of \cite{PWunstable} for the $S^7$ to all tri-Sasakian
manifolds. That the Englert solution has unstable modes agrees with the analysis of \cite{pagepopeinstab} where it is proved that the Englert solution corresponding to a solution with $\mathcal{N}>1$
is always unstable. On the other hand, the Englert solution associated to the $\mathcal{N}=1$ solution does not have such modes within the truncation, which does not necessarily mean it is stable as there still might be problematic modes not contained in the truncation. Presumably the stability depends on the concrete tri-Sasakian manifold. For instance, it is known that the Englert solution on the squashed S$^7$ is unstable \cite{itoinstab}.

\subsection{Embedding in the spectrum on $S^7$}\label{sec:ComparisonS7}

Above we studied the supermultiplet structure of the spectrum on the $\mathcal N=3$ and the $\mathcal N=1$ AdS vacua. Since $S^7$ admits a tri-Sasakian structure
it should be possible to embed the spectra computed above into the 11D supergravity spectrum on AdS$_4\times S^7$. The complete spectrum of fluctuations around the supersymmetric $\mathcal{N}=8$ solution with round $S^7$ metric can be found in table 9 of \cite{KKreview}, which also displays how the mass eigenstates transform under the $R$-symmetry group SO(8) of maximal supergravity. We find that decomposing
\eq{
{\rm SO}(8) \to {\rm Sp}(2)\times{\rm SO}(3)_R\,
}
and keeping just the modes invariant under Sp(2), a finite number of states is obtained, which moreover exactly match
the spectrum of either the $\mathcal{N}=3$ supersymmetric solution or its skew-whiffed counterpart. More precisely, which spectrum
one obtains depends on the embedding of ${\rm Sp}(2)\times{\rm SO}(3)_R$ in SO(8).
Indeed, as explained for instance in \cite[p.~79]{KKreview}, because of triality there are three different embeddings, $\text{Sp}(2)_\pm$ and $\text{Sp}(2)_v$, depending on which of the three ${\bf 8}$ representations, the positive and
negative chirality spinors ${\bf 8}_\pm$ and the vector ${\bf 8}_v$, of SO(8) gets decomposed as ${\bf 8} \to ({\bf 5}, {\bf 1}) + ({\bf 1}, {\bf 3})$.

We find that requiring invariance under $\text{Sp}(2)_\pm$ leads
to the truncations of the spectrum on $S^7$ displayed in table \ref{decompsusy} and table \ref{decompskew} respectively.
\begin{table}[tp]
\begin{center}
\rowcolors{2}{white}{gray!50}
\begin{tabular}{|c|c|l|c|}
\hline
 Field & $n$ & SO(8)$\,\to\, $SO(3)$_R$ & $\Delta$  \\
\hline
graviton & 0 & $(0,0,0,0) \to {\bf{1}}$ & $3$ \\\hline
gravitinos & 0 & $(0,0,0,1) \to {\bf 3}$ & $5/2$ \\
& 2 & $(1,0,1,0) \to {\bf 1}$ & $9/2$ \\
\hline
vectors & 0 & $(0,1,0,0) \to {\bf 3}$ & $2$ \\
& $2$ & $(1,0,1,1) \to {\bf 3}$ & $4$ \\
& $2$ & $(0,1,0,0) \to {\bf 3}$ & $5$ \\
\hline
spin 1/2 & 0 & $(1,0,1,0) \to {\bf 1}$ & $3/2$ \\
& 2 & $(1,1,1,0) \to {\bf 3}$ & $7/2$ \\
& 2 & $(0,1,0,1) \to {\bf 3}+{\bf 5}$ & $9/2$ \\
& 2 & $(0,0,0,1) \to {\bf 3}$ & $11/2$ \\
\hline
scalars & 2 & $(2,0,2,0) \to {\bf 1}$ & $3$ \\
& 2 & $(0,2,0,0) \to {\bf 1}+{\bf 5}$ & $4$ \\
& 2 & $(0,0,0,2) \to {\bf 1}+{\bf 5}$ & $5$ \\
& 2 & $(0,0,0,0) \to {\bf 1}$ & $6$ \\
\hline
\end{tabular}
\caption{Sp(2)$_+$-invariant states of the spectrum of the round $S^7$.}
\label{decompsusy}
\end{center}
\end{table}
\begin{table}[tp]
\begin{center}
\rowcolors{2}{white}{gray!50}
\begin{tabular}{|c|c|l|c|}
\hline
 Field & $n$ & SO(8)$\,\to\, $SO(3)$_R$ & $\Delta$ \\
\hline
graviton & 0 & $(0,0,0,0) \to {\bf{1}}$ & \;\; $3$ \;\;\\\hline
gravitinos & 1 & $(1,0,0,1) \to {\bf 1}$ & $3$ \\
& 1 & $(0,0,1,0) \to {\bf 3}$ & $4$ \\
\hline
vectors & 0 & $(0,1,0,0) \to {\bf 3}$ & $2$ \\
& 2 & $(1,0,1,1) \to {\bf 3}$ & $4$ \\
& 2 & $(0,1,0,0) \to {\bf 3}$ & $5$ \\
\hline
spin 1/2 & 1 & $(0,1,1,0) \to {\bf 3}+{\bf 5}$ & $3$ \\
& 3 & $(0,1,0,1) \to {\bf 3}$ & $5$ \\
& 3 & $(1,0,0,1) \to {\bf 1}$ & $6$ \\
\hline
scalars & 0 & $(0,0,2,0) \to {\bf 1}+{\bf 5}$ & $2$ \\
& 2 & $(0,2,0,0) \to {\bf 1}+{\bf 5}$ & $4$ \\
& 2 & $(0,0,0,0) \to {\bf 1}$ & $6$ \\
& 4 & $(2,0,0,2) \to {\bf 1}$ & $6$ \\
\hline
\end{tabular}
\caption{Sp(2)$_-$-invariant states of the spectrum of the round $S^7$.}
\label{decompskew}
\end{center}
\end{table}
In each case we started from the spectrum found in table 9 of \cite{KKreview}.
As in \cite{KKreview} we use the Dynkin labels to denote the different
representations of SO(8). Furthermore $n$ is the KK level, where $n=0$ corresponds to the massless $\mathcal{N}=8$ supergravity
multiplet, and we only display the representations that have Sp(2)-invariant modes. For completeness we also added the modes
with half-integer spin. In particular, the total number of Sp(2)-invariant gravitinos corresponds to the number of supersymmetries
in the theory ($\mathcal{N}=4$), while the number of massless gravitinos corresponds to the number of supersymmetries preserved by the solution (respectively $\mathcal{N}=3$ and $\mathcal{N}=0$).
As for the embedding $\text{Sp}(2)_v$ it leads to an infinite tower of states.

We find indeed that the $\text{Sp(2)}_+$-invariant spectrum matches the supersymmetric spectrum in table \ref{spectrumN3scal}, with the $\mathcal N=3$ gravity multiplet being in the lowest KK level ($n=0$), and the massive $\mathcal N=3$ gravitino multiplet coming from the second KK iteration ($n=2$). The $\text{Sp}(2)_-$-invariant spectrum on the other hand matches the skew-whiffed spectrum, with contributions up to the fourth KK level. This agrees with the fact that switching the two spinorial representations of SO(8) amounts to an inversion of the orientation or, equivalently, to flipping the sign of the four-form parameter $k$.

One obtains novel consistent truncations by restricting the massless $\mathcal{N}=8$ consistent truncation on $S^7$
to either $\text{Sp}(2)_\pm$-invariant modes. Actually, being $\text{Sp}(2)$-invariant these modes exist on every tri-Sasakian manifold and not just on $S^7$. Equivalently, one can start from the tri-Sasakian reduction and further truncate to the massless sector ($n=0$).
We will present these truncations in section \ref{sec:masslesssubtrunc}.

To conclude this section it is interesting to have another look at the instability of the Pope--Warner
solution and compare with the analysis of \cite{PWunstable}, where the modes of the massless $\mathcal{N}=8$ truncation on $S^7$
are considered that are invariant under $\text{SU(4)}_-$. In a similar way as for Sp(2), the $\text{SU(4)}_\pm$
are defined as the embeddings of $\text{SU}(4) \times \text{U}(1)_R$ in SO(8) that decompose the respective spinors
as ${\bf 8}_\pm \to {\bf 6}_0 + {\bf 1}_+ + {\bf 1}_-$. In fact: $\text{Sp}(2)_\pm \subset \text{SU}(4)_\pm$.
We found one unstable mode {\em within} our truncation with $m^2=-3$, while
in \cite{PWunstable} unstable modes are found within the massless $\mathcal{N}=8$ truncation, but transforming
as a $\bf{20'}$ under SU(4)$_-$ and thus outside the SU(4)$_-$-invariant truncation. Under the decomposition of SU(4)$_-$
under the subgroup Sp(2)$_-$ the $\bf{20'}$ has indeed exactly one singlet, which should thus be the same mode as our unstable mode.

\section{Consistency proof for universal truncations from cosets}
\label{sec:CosetRealiz}

Dimensional reduction on coset manifolds $G/H$ leads to consistent truncations. Indeed, as we mentioned in the introduction a
truncation ansatz keeping all and only the fields that are invariant under the global symmetry generated by the action of $G$ is guaranteed to be consistent. Generically, the truncation will contain massive KK modes.

In this section, we show that the universal tri-Sasakian truncation takes exactly the same form as the $\text{Sp}(2)$-invariant truncation on the Sp(2)/Sp(1) coset space. Since consistency of the latter is guaranteed by $\text{Sp}(2)$-invariance, this provides us with a simple proof of consistency of the former. We stress that this holds although generically a tri-Sasakian manifold is {\em not}
a coset manifold.

We also show that, in fact, all universal consistent truncations with massive modes worked out in the recent literature can be seen as formally equivalent to $G$-invariant truncations on coset manifolds.

Finally, we demonstrate that a $G$-invariant reduction on the coset manifold $N^{010}$ enhances the universal tri-Sasakian reduction with a vector multiplet of $\mathcal N=4$ supergravity (Betti multiplet). We discuss the generality of this phenomenon in the other universal consistent truncations.

The relations between the various consistent truncations discussed in this section are summarized in figure \ref{Diagram} below.

\subsection{Coset technology}\label{CosetTechno}

We start by briefly introducing some general concepts about coset manifolds $G/H = \{gH: g\in G \}$ (for more details see e.g.\ \cite{cosetreview1}).
Take $L(y)$ a coset representative. Let $\{ \mathcal{H}_a\}$ be a basis of generators of the algebra of $H$, and $\{ \mathcal{K}_{i}\}$ a basis of the complement of that algebra inside the algebra of $G$.
The decomposition of the Lie-algebra valued one-form
\al{
L^{-1}\d L \,=\, e^i\mathcal{K}_i+\omega^a\mathcal{H}_a
~,}
defines a coframe $e^i(y)$ on $G/H$. Moreover, in terms of the structure constants of $G$ one finds
\eq{\label{dcommut}
\d e^i \,=\, -\frac{1}{2}f^i{}_{jk}e^j\wedge e^k-f^i{}_{aj}\omega^a\wedge e^j \, .
}
We are interested in forms that are invariant under the left action
of $G$ on $G/H$, called {\em left-invariant} forms. One can show that an $l$-form $\phi$ is left-invariant if and only if it can be written as
\eq{
\phi \,=\, \frac{1}{l!}\phi_{i_1\dots i_l}e^{i_1}\wedge\dots \wedge e^{i_l} \, ,
}
with the components $\phi_{i_1\dots i_l}$ being independent of the coset coordinates $y$, and satisfying
\eq{\label{leftinv}
f^j{}_{a[i_1}\phi_{i_2\dots i_l]j}=0~.
}
Upon taking the exterior derivative, $\d \phi$, condition \eqref{leftinv} ensures
that the part coming from the second term in \eqref{dcommut} drops, which implies that
the exterior derivative preserves the left-invariance property.
The left-invariance condition for a metric, $\d s^2_{G/H}=g_{ij} e^i e^j$, is analogous, namely $g_{ij}$ has to be independent of $y$ and needs to satisfy
\eq{
f^k{}_{a(i} \,g_{j)k} = 0 \, .
}
As an aside one can show that
harmonic forms must be left-invariant and thus the cohomology of the coset manifold
is isomorphic to the cohomology of left-invariant forms.

Another important property for our purposes is that coset spaces $G/H$ are particularly simple examples of $H$-structure manifolds. Indeed, one can prove that if $H$ contains no nontrivial
invariant subgroup of $G$, then the structure group of the tangent bundle on the coset $G/H$ can be reduced to $H$ (see e.g.\ \cite[app.\ A]{cosets}). To characterize the $H$-structure one can use the set of left-invariant forms, which by definition are invariant under the local action of $H$ (this is condition \eqref{leftinv}). Moreover, since the exterior derivative of a left-invariant form is still left-invariant, the torsion associated with the $H$-structure is also $H$-invariant, and constant.

\subsection{Equivalence between tri-Sasakian and $\text{Sp}(2)/\text{Sp}(1)$ reduction}
\label{sec:Sp2Sp1}

Upon comparing the spectrum of the tri-Sasakian reduction with the one of 11D supergravity on AdS$_4\times S^7$,
we found in section \ref{sec:ComparisonS7} that the tri-Sasakian truncation preserves precisely the modes that are invariant under an Sp(2) subgroup of the SO(8) symmetry group. Sp(2) is known to be the smallest simply-connected subgroup of SO(8) having a transitive action on $S^7$, and its isotropy subgroup is Sp(1). It follows that there exists a particular description of $S^7$ as the 7D coset manifold Sp(2)/Sp(1)$_S$. This describes both the round and squashed seven-sphere \cite{casromanssym}, and captures precisely its tri-Sasakian structure, with $\text{Sp}(1)_S \simeq \text{SU}(2)_S$ being the structure group introduced in section~\ref{sec:3Sprelim}.

In the following we construct a tri-Sasakian structure in terms of left-invariant forms.
This will allow us to show that a left-invariant truncation on Sp(2)/Sp(1)$_S$ is equivalent in form to the universal tri-Sasakian reduction. Consider
\eq{
\label{Sp2Sp1}
\frac{G}{H}\, =\, \frac{\text{Sp(2)}}{\text{Sp(1)}} \, ,
}
where the $\text{Sp}(1)\simeq \text{SU}(2)$ is embedded in such a way that it corresponds
to one of the two factors in a $\text{SU(2)} \times \text{SU}(2) \simeq \text{Spin}(4)$ subgroup of $\text{Sp}(2) \simeq \text{Spin}(5)$. Among the different possible embeddings of Sp(1) in Sp(2), this is the one describing the seven-sphere \cite[pp.$\:$41, 42]{KKreview}.

For the generators we make the following convenient choice:
\eq{\spl{
& (\mathcal{K}_i) \,=\, \Big(t_{15},\, t_{25},\, t_{35},\, t_{45},\, \frac{1}{\sqrt{2}} (t_{23} + t_{14}),\,\frac{1}{\sqrt{2}} (t_{31} + t_{24}),\,\frac{1}{\sqrt{2}} (t_{12} + t_{34})\Big) \, , \\
& (\mathcal{H}_a) \,=\, \Big(\frac{1}{\sqrt{2}} (t_{23} - t_{14}),\,\frac{1}{\sqrt{2}} (t_{31} - t_{24}),\,\frac{1}{\sqrt{2}} (t_{12} - t_{34})\Big) \, ,
}}
where $(t_{mn})^{pq} = 2 \, \delta_{[m}^{p}\delta_{n]}^q $ are so(5) generators.
Using the formulae in section \ref{CosetTechno} we construct the complete set of left-invariant forms. We find that the
one-forms and two-forms are spanned by
\eq{\label{3SformsSp2Sp1}\spl{
\eta^I & \,=\, \frac{1}{\sqrt{2}} \{ e^5,\, e^6, \,e^7 \} \, , \\
J^I & \,=\, \frac{1}{4 }\{ -e^1 \wedge e^4 - e^2 \wedge e^3,\, e^1 \wedge e^3 - e^2 \wedge e^4,\, -e^1 \wedge e^2 - e^3 \wedge e^4 \} \, ,
}}
which satisfy the eqs.~\eqref{dereta}, \eqref{derJ} and \eqref{wedgeJ} and therefore provide a tri-Sasakian structure. Furthermore, the only left-invariant three-forms are wedges of the above one- and two-forms. Likewise, the general left-invariant metric takes
the form of the 7D part of the tri-Sasakian ansatz \eqref{metricansatz} (with $\d s^2(B_{\text{QK}})=(e^1/2)^2 + \cdots+(e^4/2)^2$).

$\text{Sp}(2)/\text{Sp}(1)$ is therefore a {\em minimal} coset realization of the tri-Sasakian geometry. Upon using eq.~\eqref{3SformsSp2Sp1} the reduction ansatz based on left-invariant forms looks exactly the same as the tri-Sasakian reduction
ansatz. Moreover, the differential properties are the same so that one ends up with the same 4D equations of motion.
Hence the consistency of the universal reduction on a tri-Sasakian manifold
is derived in this way from the consistency of the reduction on the coset manifold Sp(2)/Sp(1).

We can also illustrate the origin of SO(3)$_R$ in the coset geometry. In \cite{casromanssym} it is explained that the isometry group
of a coset $G/H$ is given by $G \times N(H)/H$, where the $G$ acts on the left and $N(H)/H$ from the right. $N(H)$ is the normalizer, defined as $N(H):=\{g\in G: gH = Hg\}$. Moreover only $N(H)/H$ corresponds to left-invariant isometries.
In the case of the coset $\text{Sp}(2)/\text{Sp}(1)$, we find $N(H)/H = \text{SO}(3)$. The total isometry group $\text{Sp}(2) \times \text{SO}(3)$ is to be identified with the subgroup of SO(8) discussed in section \ref{sec:ComparisonS7} (which embedding we choose is not important at this stage because we have not fixed the sign of $k$ in the four-form). The fields in the coset reduction are invariant under the first factor and can transform non-trivially under the second factor. Indeed, the generators of the latter can be identified with the tri-Sasakian Killing vectors $\xi_I$, hence this is precisely the SO(3)$_R$ that is gauged in the reduction. Also
from the point of view of the dual CFT theory it is worth keeping in mind that we are describing a sector and possible deformations
that are invariant under the Sp(2) subgroup of the full SO(8) global symmetry group of ABJM \cite{abjm}.

\subsection{Coset realizations of the other universal truncations}

The equivalence between universal truncations and coset reductions goes beyond the tri-Sasakian case discussed above: it can actually be extended to all universal consistent truncations that recently appeared in the literature. To see this, let us consider again the seven-sphere, which is known to admit the following coset descriptions:
\eq{ S^7\;: \quad \frac{{\rm Sp}(2)}{{\rm Sp(1)}}\,,\quad \frac{{\rm SU}(4)}{{\rm SU(3)}}\,, \quad \frac{{\rm Spin}(7)}{{\rm G_2}}\,,\quad \frac{{\rm SO}(8)}{{\rm SO(7)}}\,.
}
These preserve more and more isometries,
\eq{
\label{groupchain}
\text{Sp}(2) \subset \text{SU}(4) \subset \text{Spin}(7) \subset \text{SO}(8) \, ,
}
and therefore allow for less and less left-invariant deformations and lead to more and more generic structure group.
Indeed, as explained in section \ref{CosetTechno}, these cosets $G/H$ define simple $H$-structures, where $H$ is
Sp(1)$\,\simeq\,$SU(2), SU(3), G$_2$ and SO(7), respectively. While the first case gives the tri-Sasakian structure studied above,
the reduction on left-invariant forms on Spin(7)/G$_2$ and SU(4)/SU(3) provides minimal coset representatives for respectively the weak G$_2$ and the Sasaki--Einstein reduction. As a check, weak G$_2$ and Sasaki--Einstein structures are indeed simple examples of G$_2$- and SU(3)-structures with invariant and constant torsion. The last case
leads to a reduction ansatz containing only a 4D metric and a breathing mode, which can be extended to any Einstein manifold.
We conclude that a left-invariant truncation on these coset manifolds is equivalent in form to the universal $\mathcal N=2$, $\mathcal N=1$ and $\mathcal N=0$ truncations studied in \cite{vargaun2}.

It follows from eq.~\eqref{groupchain} that the truncations of \cite{vargaun2} can be seen as subtruncations of the tri-Sasakian reduction. We saw in section \ref{sec:3Sprelim} that a tri-Sasakian structure admits an $S^2$ worth of Sasaki--Einstein structures: by picking one of these, we can further truncate our reduction ansatz in such a way as to reproduce the reduction ansatz of \cite{vargaun2}.
Choosing for instance $\alpha^I = \delta^I_3$ in \eqref{subsasaki}, so that the surviving U(1) is generated by $\xi_3$, we obtain a
Sasaki--Einstein subtruncation by setting
\begin{equation}\label{sasakitrunc}
\begin{array}{rlll}
& V_3 = V \, , &V_1 = V_2 = U \, , \quad\qquad \qquad & A_1^3=A_1 \, , \\
&(c_2)_3 = c_2 \, , &(c_1)_3 = c_1 \, , \qquad\qquad &(\tilde{c}_1)_3 = - c_1 \, , \\
&c_{11} = -c_{22} = \Re c_{\Omega} \, ,\qquad\qquad &c_{12} =c_{21}=\Im c_{\Omega} \, ,& \\
&c_{33} = c_0 \, ,  &\chi  = - c_0 \, ,
\end{array}
\end{equation}
with the other matter fields vanishing. Next, as discussed in \cite{vargaun2}, one can successively truncate down to $\mathcal N=1$ and $\mathcal N=0$ by going from the Sasaki--Einstein ansatz over the weak G$_2$ ansatz to the Einstein ansatz. Hence starting from the tri-Sasakian reduction we obtain a sequence of subtruncations with decreasing amount of supersymmetry on geometries with increasing structure group.

More instances of universal supersymmetric consistent truncations on simple $H$-structure manifolds have been constructed, and again they turn out to be equivalent to left-invariant reductions on a coset manifold. One example is given by massive type IIA supergravity on nearly-K\"ahler manifolds, whose study of consistency was initiated in \cite{poorNK}.
This reduction, which leads to an $\mathcal N=2$ gauged supergravity model \cite{ExploitingN=2}, turns out to be
formally equivalent to a reduction on the 6D coset manifold G$_2$/SU(3). Yet another example is the universal $\mathcal N=4$ reduction of type IIB supergravity on a five-dimensional Sasaki--Einstein manifold \cite{IIBonSE,vargaun3}, which is reproduced by a reduction on the SU(3)/SU(2) coset~\cite{IIBonSE}.

\begin{figure}
	\centering
\includegraphics[width=15cm]{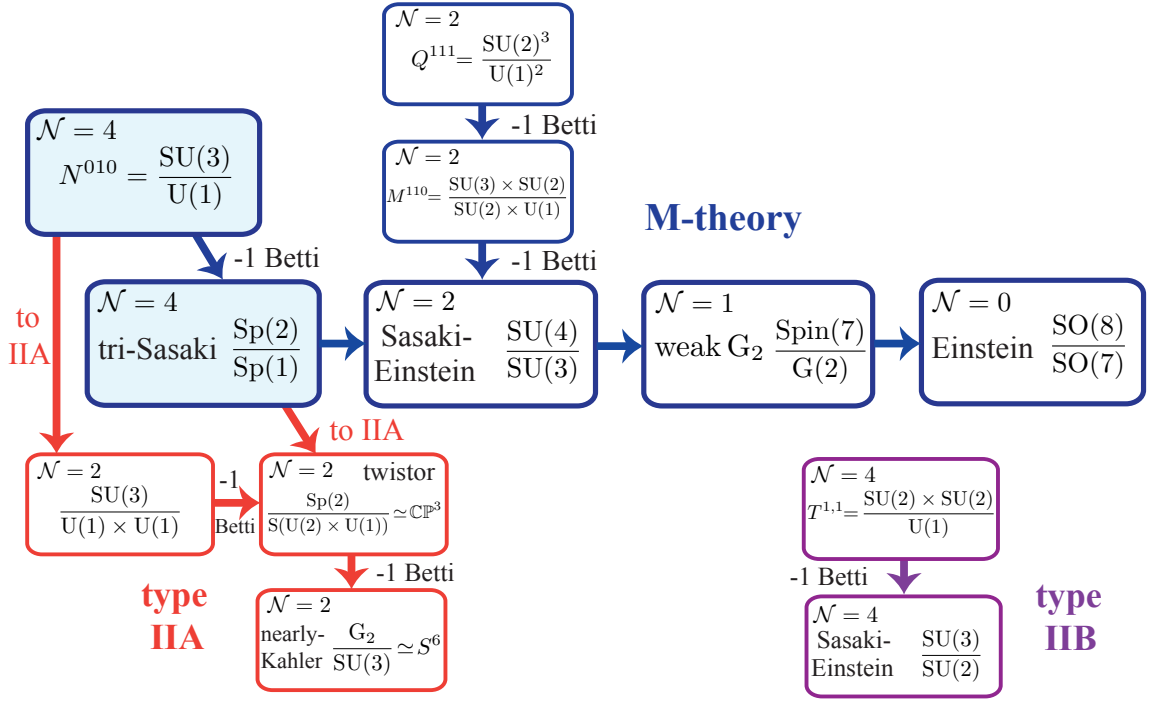}
\caption{\small Web of interrelated consistent truncations of 11D, type IIA and type IIB supergravity. In each box we display the geometric structure (if any) and the coset manifold on which the truncation is based, as well as its amount of supersymmetry. Each arrow denotes a consistent subtruncation. The colored boxes highlight the truncations on which we focused in this paper.}
\label{Diagram}
\end{figure}

\subsection{An enhanced $\mathcal N=4$ truncation on the coset $N^{010}$}
\label{sec:N010}

Besides $\text{Sp}(2)/\text{Sp}(1)\simeq S^7$, there is another homogeneous representative of the class of 7D tri-Sasakian manifolds, namely the coset space $N^{010}$. In the following we show that a left-invariant truncation on this space is not minimal in that
it enhances the universal tri-Sasakian reduction with an $\mathcal N=4$ vector multiplet.

Consider the family of coset manifolds $N^{pqr}$, defined as \cite{casnpqr}
\eq{
\frac{\text{SU(3)}\times\text{U(1)}}{\text{U(1)}\times\text{U(1)}} \, ,
}
where the U(1) {\em not} divided out is given by $Z=\frac{1}{2i}(p \sqrt{3} \lambda_8 + q \lambda_3 + r T)$. Here $\lambda_i$
are the Gell-Mann matrices generating SU(3) and $T$ is the generator of the U(1) in the numerator. For definiteness,
for each $p$ and $q$ we restrict to the manifolds $N^{pq0}$, which are the universal covers within that family. The U(1) in the
numerator is then completely quotiented out so that we end up with
\eq{\label{Npq0coset}
\frac{G}{H}\, =\, \frac{\text{SU(3)}}{\text{U(1)}} \, .
}
The generators are chosen as:
\eq{\spl{
& (\mathcal{K}_i) \,=\, \frac{1}{2i} \Big(\lambda_1,\lambda_2,\lambda_4,\lambda_5,\lambda_6,\lambda_7,\sqrt{3}p \lambda_8+ q \lambda_3\Big) \, , \\
& (\mathcal{H}_a) \,=\, \frac{1}{2i} \Big(-\sqrt{3}q \lambda_8+3p \lambda_3\Big) \, ,
}}
and the structure constants of SU(3) are split accordingly. The topology depends only on $x=3p/q$, so we can take $p$
and $q$ to be relatively prime integers. Furthermore, there is a symmetry of the structure constants which can be used to put
\subeq{\al{
& p \longrightarrow p \, , \qquad q \longrightarrow -q \, , \\
& p \longrightarrow - (p+q) \, \qquad q \longrightarrow q-3p \, ,
}}
so that the corresponding $N^{pq0}$ manifolds are topologically equivalent, and which can be used to take $0 \le x \le 1$.
It can be shown that $N^{pq0}$ is tri-Sasakian if and only if $p=0,q=1$ (and equivalent configurations).
We can construct the left-invariant, tri-Sasakian structure as follows:
\eq{\label{3SformsN010}\spl{
\eta^I & \,=\, \frac{1}{2} \{ e^1,\, e^2,\, e^7 \} \, , \\
J^I & \,=\, \frac{1}{8 }\{ -e^3 \wedge e^6 + e^4 \wedge e^5,\, -e^3 \wedge e^5 - e^4 \wedge e^6,\, -e^3 \wedge e^4 + e^5 \wedge e^6 \} \, .
}}
However, in contrast to the case of Sp(2)/Sp(1), the tri-Sasakian forms in \eqref{3SformsN010} are not the only forms left-invariant under the action of $G=\,$SU(3).
There are no extra one-forms, but there is an extra left-invariant two-form,
\eq{
\Phi = e^{34} + e^{56} \, ,
}
which is closed, $\d \Phi =0$, and not exact, reflecting the non-trivial cohomology of $N^{010}$.
 This implies that there are also three extra left-invariant three-forms of the type $\eta^I \wedge \Phi$, and three new left-invariant
symmetric two-tensors, which can be used in the reduction ansatz for the metric,
\eq{
\{ (e^3)^2 + (e^4)^2 - (e^5)^2 - (e^6)^2,\,\, e^3 e^5 + e^4 e^6,\,\, e^3 e^6 - e^4 e^5 \} \, .
}
In all, a reduction ansatz based on left-invariant metric and forms on the coset $N^{010}$ leads to six extra scalars and one extra vector compared to the universal tri-Sasakian reduction ansatz. These combine into an extra $\mathcal N=4$ vector multiplet. Extra multiplets appearing due to the non-trivial cohomology of the compactification manifold are sometimes dubbed Betti multiplets \cite{DAuriaFre_BoseFermi}. We remark that this extra multiplet is not contained in the spectrum of $S^7$. The full spectrum of 11D supergravity on AdS$_4 \times N^{010}$ and its $\mathcal N=3$ supermultiplet structure was studied in \cite{Termonia:1999cs, Fre':1999xp}.

It is interesting to notice that the left-action of $G=\,$SU(3) should correspond to a global symmetry in the dual $\mathcal N=3$ conformal field theory. So the left-invariant reduction amounts in the dual theory to truncating to the singlet sector of this symmetry.

We remark that the phenomenon of enhancement with Betti multiplets is common in the other universal truncations mentioned in the previous subsection. For the truncation based on 7D Sasaki--Einstein structures, one or two extra $\mathcal N=2$ Betti vector multiplets are obtained if one performs a left-invariant reduction on the cosets $M^{110}$ and $Q^{111}$, respectively. The universal type IIB reduction on 5D Sasaki--Einstein structures \cite{IIBonSE, vargaun3} is enhanced with an $\mathcal N=4$ multiplet if one considers the $T^{1,1}$ coset~\cite{T11red}. In this case, the $S^2\times S^3$ topology of $T^{1,1}$ allows to also introduce three-form fluxes, and the extra fields become crucial for many gauge-gravity applications based on the conifold geometry.
The nearly-K\"ahler (or, equivalently, G$_2$/SU(3)) reduction of massive type IIA supergravity gets enhanced with one or two $\mathcal N=2$ Betti vector multiplets if a left-invariant truncation is performed on the 6D coset manifold Sp(2)/S(U(2)$\times$U(1)) or SU(3)/(U(1)$\times$U(1)), respectively \cite{ExploitingN=2}.

As a final comment, we observe that, in the limit of vanishing Romans mass, the two latter truncations are subtruncations of the universal tri-Sasakian (or, equivalently, Sp(2)/Sp(1)) reduction and of the $N^{010}$ reduction, respectively. This can be seen by performing a circle reduction of the 11D reduction ansatz, choosing the U(1) generated by one of the Killing vectors $\xi_I$. In this process only the fields that are invariant under the chosen U(1) are kept, and the 4D spectrum gets truncated in such a way that the aforementioned type IIA reductions are obtained.
These $\mathcal N=2$ gauged supergravity models admit a further gauging corresponding to the introduction of the Romans mass in type IIA supergravity. It would be interesting to study whether there exists a deformation of our $\mathcal N=4$ action which can reproduce upon U(1) reduction these massive type IIA models. This gauging would involve the vectors $\tilde A_{1}^I$ being magnetic partners of the $A_1^I$ coming from the 11D metric and coupling to the KK monopole.

\section{New subtruncations}\label{sec:subtruncations}

In addition to the ones described in the previous section, our $\mathcal N=4$ tri-Sasakian reduction admits further simple consistent subtruncations, that contain many less degrees of freedom and should be useful for several gauge-gravity applications.
In particular, two of them still contain a non-abelian gauge group SO(3), which makes them an interesting playground for embedding into M-theory holographic condensed matter phenomena with vector order parameter, like $p$-wave superconductors \cite{Gubser:2008zu,GubserPufu}.

\subsection{Truncation to minimal $\mathcal N=3$ gauged supergravity}
\label{sec:masslesssubtrunc}

The first case is the consistent subtruncation to minimal $\mathcal N=3$ gauged supergravity. This is obtained by projecting out the massive $\mathcal N=3$ gravitino multiplet discussed in section \ref{sec:spectrum}.
Concretely, we fix $k =+1$ and set to zero all the scalar fields, as well as the massive vector combinations of table \ref{spectrumvec} (now promoted to non-linear order). So we put:
\begin{equation}
3 A_1^I - (a_1^I -2 \tilde{a}_1^I) = 0 \,,\qquad
c_{1I} + \tilde{c}_{1I} = 0\,.
\end{equation}
Then for the massless combination we can take just $A_1^I$. Plugging this in the duality relations of appendix \ref{app:dualityrel} we obtain the further relations
\eq{\label{minimalN=3}\mathcal Da_{1I} = -\mathcal D\tilde a_{1I} = -*\mathcal D c_1 = *\mathcal D \tilde c_1 = F_{2I}\,.}
With these constraints, all the equations of motion following from the $\mathcal N=4$ action given in section \ref{sec:resultaction} are automatically satisfied, except those for the metric and the SO(3)$_R$ gauge fields $A_1^I\,$. These follow from the truncated action
\eq{S = \frac{1}{2\kappa_4^2}\int \left(R - 2 F_{2I}\cdot F_2^I +24\right)*1\,.}
We conclude that on every 7D tri-Sasakian manifold one can consistently truncate 11D supergravity to minimal $\mathcal N=3$ gauged supergravity in four dimensions.\footnote{The original reference for minimal $\mathcal N=3$ SO(3) gauged supergravity is \cite{FreedmanDas}. To recover the lagrangian given therein (with $e =-1$), one needs to rescale $g_{\mu\nu}\to g_{\mu\nu}/4$ and $A_1^I \to A_1^I/2$.} This represents a further example in support of the conjecture made in \cite{vargaun1} about the existence of consistent truncations to minimal gauged supergravity associated with any supersymmetric AdS solution of higher-dimensional supergravity.

Let us make a final comment regarding the specific instance of $S^7$ (note that for this case the truncation above was already described in \cite{Maldacena:2004rf}). Recalling the discussion of section \ref{sec:ComparisonS7}, we observe that the AdS spectrum of the present truncation is given by the intersection between the lowest ($n=0$) KK level and the KK modes invariant under Sp(2)$_+$. Hence the action (\ref{minimalN=3}) describes precisely the sector of our $\mathcal N=4$ truncation that is also contained in the maximal SO(8) gauged supergravity arising from the $S^7$ reduction of \cite{deWitS7} (which keeps the modes from the lowest KK level), and consistency is ensured.

It is also possible to truncate to the Sp$(2)_-$-invariant sector of maximal SO(8) gauged supergravity. Besides the metric, the preserved degrees of freedom are given by the $n=0$ sector of table \ref{decompskew}: we have three gauge bosons and six scalars.
Since no fields with half-integer spin in the $n=0$ sector are invariant under Sp$(2)_-$, this is a purely bosonic truncation.

\subsection{$\mathcal N=1$ truncation to the SO(3)$_R$-invariant sector}

There is also a truncation in which we switch off all the fields charged under SO(3)$_R$. The surviving bosonic fields are just the metric and four scalars:
\eq{U\,,\quad V \equiv V_1=V_2=V_3\,,\quad \chi\,,\quad  c \equiv c_{11} = c_{22} = c_{33}\,.}
The truncated action is
\eq{\label{N=1action}\spl{
S = \frac{1}{2\kappa_4^2} \int &\Big[R_4 - 12 (\d U)^2 - 12 \, \d U\! \cdot\! \d V
- \frac{15}{2} (\d V)^2   - 3\,e^{-4U-2V}(\d c)^2  - \frac 12 \, e^{-6 V} (\d\chi )^2\\
&+\,6\, e^{-4U -5V} +48 \, e^{-6U -3V}  -12 \, e^{-8U - V}
-\, 72 \, e^{-12U -3V}  c^2 \\
&-\, 12 \, e^{-8U -7V}
 (\chi  +\, c)^2 -  18\,e^{- 12 U - 9V}( c^2 + 2\, \chi \, c - k)^2\Big]*1\,.}}

Of the original $\mathcal N=4$ gravitini, just one survives the truncation. In a sense, this truncation is complementary to the one presented in the previous subsection, because here we truncate the three supersymmetries of $\mathcal N=4$ that were preserved there, while the supersymmetry truncated there is the one preserved here. A first clue that this truncation should be supersymmetric
can be found from table \ref{spectrumN1scal}, where we see that the SO(3)$_R$-invariant scalar fluctuations organize in supermultiplets around the $\mathcal N=1$ vacuum (the conformal dimensions differ by one). Furthermore, one can show that this action
takes the standard $\mathcal{N}=1$ supersymmetric form with the following K\"ahler potential and superpotential\footnote{See \cite{micusuppot} for the expression of the superpotential for a general compactification of M-theory on a weak G$_2$-manifold.}
\eq{
\mathcal{K} = - \ln [-i(\tau - \bar{\tau})] -6 \, \ln [-i (z - \bar{z})]  \, , \qquad \mathcal{W}= 12 \sqrt{2} \, [z(z+ 2 \,\tau) - k] \, ,
}
where we defined the complex fields
\eq{
\tau = \chi + i \,e^{3 V} \, , \qquad  z = c + i\, e^{2 U + V}  \, .
}
One can also check that for $k<0$ the squashed AdS solution \eqref{squashedSol} satisfies the standard $\mathcal N=1$ supersymmetry condition $(\partial + \partial K)\mathcal W =0$.

It is consistent to further truncate the scalars $\chi$ and $c$, which breaks supersymmetry and for the specific case of $S^7$ leads to a model studied long ago in \cite{Page:1984qv}.

Since it contains all the SO(3)$_R$-invariant AdS$_4$ solutions presented in section \ref{sec:vacsol}, it would be interesting to exploit this constrained setup to study domain walls interpolating between them. In particular, a domain wall connecting the round \eqref{3Ssol} and the squashed \eqref{squashedSol} solutions would describe a holographic renormalization group flow between the (skew-whiffed, i.e.\ $\mathcal N=0$) ABJM theory \cite{abjm} and the $\mathcal N=1$ theory of~\cite{oogurisquashed}.

We will come back to applications of these consistent truncations in the near future.

\bigskip
\acknowledgments

We thank Gianguido Dall'Agata, Anton F.\ Faedo, Dario Martelli and Peter West for useful conversations.
The work of D.C. was supported in part by the Fondazione Cariparo Excellence Grant {\em String-derived supergravities with branes and fluxes and their phenomenological implications}. P.K.\ is a Postdoctoral Fellow of the FWO -- Vlaanderen. The work of
P.K.\ is further supported in part by the FWO -- Vlaanderen project
G.0651.11 and in part by the Federal Office for Scientific,
Technical and Cultural Affairs through the `Interuniversity
Attraction Poles Programme Belgian Science Policy' P6/11-P.

\appendix

\section{Conventions}
\label{sec:conventions}

The Hodge dual of an $l$-form $\phi$ in $D$ dimensions is a $(D-l)$-form given by
\eq{
(* \phi)_{\mu_1\ldots\mu_{D-l}} = \frac{1}{l!} \sqrt{|\det g|} \epsilon_{\mu_1 \ldots \mu_{D-l}\nu_1 \ldots \nu_l} g^{\nu_1\rho_1} \cdots g^{\nu_l\rho_l} \phi_{\rho_1 \ldots \rho_l} \, ,
}
where $g$ is the $D$-dimensional metric and $\epsilon$ the totally antisymmetric pseudo-tensor with $\epsilon_{01\ldots (D-1)}=1$ (note that
this sign choice for $\epsilon$ is opposite to \cite{SchonWeidner}). Furthermore it will be convenient to introduce the following
shorthand notation for the multiplication of forms, contracting all the indices with a weight factor:
\eq{
\label{formprod}
\phi_1 \cdot \phi_2 = \frac{1}{l!} \phi_{1,\mu_1 \ldots \mu_l} \phi_2^{\mu_1 \ldots \mu_l} = \frac{1}{l!} \phi_{1,\mu_1 \ldots \mu_l} \phi_{2,\nu_1 \ldots \nu_l} g^{\mu_1\nu_1} \cdots g^{\mu_l\nu_l} \, ,
}
and for the square:
\eq{
(\phi)^2 = \phi \cdot \phi \, .
}
With these conventions we have in particular:
\eq{
\int \phi_1 \wedge * \phi_2 = (-1)^{l(D-l)} \int \, \phi_1 \cdot \phi_2\,*1 \, .
}

\section{Auxiliary calculations for the reduction of the form sector}\label{sec:AuxCalc}

\subsection{Bianchi identities and equations of motions for the $c$-fields}
\label{app:eombiflux}

In this appendix, we derive the Bianchi identities and the second-order equations of motion for the $c$-fields.
The reduction of (\ref{BianchiG4}) gives
\subeq{\label{BianchiH}\al{
& D H_{1IJ} - 2\big(H_{2K} + \tilde H_{2K}\big)\epsilon^K{}_{IJ} + 2 \left[\left(\chi + {\rm tr}\, c\right)\delta_{JK} - 2c_{(JK)}\right]\epsilon^K{}_{LI}F_2^L = 0 \, , \\
& D H_{2I} - 2H_{3I} - H_{1JI}\wedge F_2^J = 0\, ,\\
& D \tilde H_{2I} + 2H_{3I} - \d \chi \wedge F_{2I} = 0 \, , \\
& D H_{3I} + \epsilon_{IJK}\tilde H_{2}^{\,J}\wedge F_2^K = 0\,.
}}
These are Bianchi identities in the $c$-frame, automatically solved upon plugging in the $c$-fields through the relations (\ref{Hfield-strengths}).

Reducing (\ref{EoMG4}) we get the following 4D second-order equations of motion for the $c$-fields:
\be\label{eqH4}
\d \left[e^{3(4U+V_1+V_2+V_3)}*H_4 - 4 \,c^{IJ}c_{(IJ)} + 2({\rm tr}\, c)^2 + 4\,\chi\, ({\rm tr}\, c) \right] \,=\, 0 \, ,
\ee
which is the equation for $c_3$ and is a total derivative,
\bea
&&D\left[e^{2(4U+V_1+V_2+V_3)}g^{IJ}*H_{3J}\right] \,+\, e^{3(4U+V_1+V_2+V_3)}F_2^I *H_4 \nn \\ [2mm]
&& - \, 2 \, e^{4U-V_1-V_2-V_3}\delta^{IK}g_{KJ} *\tilde H_2^J \,+\, 4 \, e^{V_1+V_2+V_3} * H_2^I \label{EoMc2}\\ [2mm]
&& +\, 4({\rm tr}\, c) \tilde H_2^I + 4 \left[ (\chi + {\rm tr}\,c)\delta^{IJ} -2\,c^{(IJ)} \,\right] H_{2J} - \epsilon^{IJK}H_{1JL}\wedge H_{1K}{}^L \,=\, 0 \, , \qquad\nn
\eea
which is the equation for $c_{2I}\,$,
\bea
&& D \left[e^{4U-V_1-V_2-V_3} g_{IJ}* \tilde H_{2}^J\right] - e^{2(4U+V_1+V_2+V_3)}\epsilon_{IJL}\, g^{JK}*H_{3K} \wedge F_2^L\nn \\[2mm]
&& -\,4\, e^{-4U} \epsilon_{IJK}\, g^{JL} * H_{1L}{}^K \,+\, 4 ({\rm tr}\,c) H_{3I} \,+\, 2 H_{1IJ}\wedge H_2^J \,=\,0 \, , \label{EoMtildec1}
\eea
which is the equation for $\tilde c_{1I}\,$,
\bea
&& D\left[e^{V_1+V_2+V_3} *H_{2I}\right] \,+\,  2\,e^{-4U} \epsilon_{IJK} \, g^{KL}* H_{1L}{}^J \, +\, \d \chi  \wedge H_{2I}\nn \\ [2mm]
&& +\, H_{1JI}\wedge \tilde H_{2}^J \,+\, 2 \left[ (\chi + {\rm tr}\,c)\delta_{IJ} -2\,c_{(IJ)} \,\right]H_3^J \,=\,0 \, , \label{EoMc1}
\eea
which is the equation for $c_{1I}\,$,
\bea
&& \d \!\left[e^{-2V_1-2V_2-2V_3}*\d \chi \right] +\,8\,e^{-8U-3V_1-3V_2-3V_3}g_{IJ}\left[ (\chi + {\rm tr}\,c)\delta^{IJ} -2\,c^{(IJ)} \,\right]*1 \nn \\ [2mm]
&& +\, e^{4U-V_1-V_2-V_3}g_{IJ} F_2^I \wedge *\tilde H_{2}^J + H_{2I}\wedge H_2^I + 4({\rm tr}\,c) H_4 = 0 \, , \label{EoMtildec}
\eea
which is the equation for the scalar $\chi\,$, and finally
\bea
&& D\left[e^{-4U} g^{IK}\delta^{JL} * H_{1KL}\right] + e^{V_1+V_2+V_3} F_{2}^I \wedge * H_{2}^J + 8\,e^{-12U-V_1-V_2-V_3}({\rm tr}\, c)\delta^{IJ} *1 \nn \\ [2mm]
&& +\, 4\,e^{-8U-3V_1-3V_2-3V_3} ( \delta^{IJ} \delta^L_K -\delta^{I}_K \delta^{JL} -\delta^{J}_K \delta^{IL} ) g_{LM}\!\left[ (\chi + {\rm tr}\,c)\delta^{KM} -2\,c^{(KM)} \,\right] *1 \nn \\[2mm]
&& +\, \tilde H_{2}^I \wedge H_{2}^J -\epsilon^{IKL} H_{3K}\wedge H_{1L}{}^J + 2 \left[ (\chi + {\rm tr}\,c)\delta^{IJ} -2\,c^{(IJ)} \,\right] H_4 \,=\,0 \, ,
\label{EoMcIJ}\eea
which is the equation of the scalars $c_{IJ}\,$.

\subsection{Duality relations}
\label{app:dualityrel}

In this appendix we construct the duality relations between the $c$-fields and the $a$-fields by
combining the 11D eqs.~\eqref{defpotC3} and \eqref{defpotC6} and reducing them.

We are only interested in the duals of $c_{2I}$, $\tilde c_{1}^{\,I}$ and $c_{1I}$ since we do not need to dualize the $c$-scalars. It is therefore sufficient to
keep the terms in $\d \widetilde C_6$ of space-time degree lower or equal to 2. We will also introduce a flux threading the 7D internal manifold
\eq{
\d \widetilde{C}_6 \rightarrow G_{7,\text{flux}} + \d \widetilde{C}_6\,,
}
with
\eq{
\label{kflux}
G_{7,\text{flux}} = 6 k \, {\rm vol_{QK}} \wedge \eta^1 \wedge \eta^2 \wedge \eta^3 \, .
}
Indeed, from the equation of motion \eqref{EoMG4} it follows that the left-hand side of \eqref{defpotC6} is closed, but it can still be non-exact, so that
$\widetilde{C}_6$ would not be globally-defined. The $\widetilde{C}_6$ introduced in \eqref{C6pot} on the other hand is globally-defined, hence a flux term is to be added separately by allowing for the closed, non-exact $G_{7,\text{flux}}$.\footnote{The normalization is chosen such that we have the $\mathcal N=3$ AdS$_4$ vacuum solution at the origin of the scalar manifold for $k=1$ (see solution \ref{N3susysol} in section \ref{sec:vacsol}).}

Proceeding we find the following duality relations between the 4D fields:
\bea
\,e^{ 3(4 U + V_1 + V_2 + V_3)} *H_4 &=& 2\left[ 2\,c^{IJ}c_{(IJ)} - ({\rm tr}\,c)^2 -2\,\chi\,({\rm tr}\,c) + 3k\right] \, , \nn \\ [2mm]
 \,e^{ 2(4 U + V_1 + V_2 + V_3)}g^{IJ}  *H_{3J} &=&  Da^I + 2a_{1}^I - 4\tilde a_{1}^I - 6k A^I + 4 c^{(IJ)}c_{1J} - 2(\chi + {\rm tr}\,c) c_{1}^I  \nn \\ [2mm]
&& -\, 2({\rm tr}\, c) \,\tilde c_{1}^I  + c_{JL}H_{1K}{}^L \epsilon^{IJK} \, , \nn \\[2mm]
\,e^{4U-V_1-V_2-V_3}g_{IJ} *\tilde H_{2}^J &=&   Da_{1I}+\epsilon_{IJK}\left( a^J F_2^K + 4\, a_2^{JK} + 3k\, A^J\wedge A^K\right)  \nn \\ [2mm]
&& -\, 2({\rm tr}\, c) \, c_{2I} + H_{1IJ}\wedge c_1^J   - c_{IJ} H_2^J \, , \nn \\ [2mm]
\, e^{V_1+V_2+V_3} *H_{2I} &=&  D\tilde a_{1I} + 2\,\epsilon_{IJK}a_{2}^{JK} + 2\, c_{(IJ)}c_2^J - (\chi + {\rm tr}\,c) c_{2I}\nn \\ [2mm]
&& -\, \frac{1}{2}\big( \tilde c_{1J}\wedge H^J_{1\;I} +  c_{1I} \wedge\d \chi  + \chi \,H_{2I} + c_{JI} \tilde H_2^J \big)\,.\nn \\
\eea
The SO(3)$_R$ covariant derivatives $D$ of the $a$-fields are defined in the same fashion as in (\ref{ccov}).
These relations can be simplified by making the following field redefinitions, which eliminate derivatives of the scalars in the duality relations and make the gauge structure of the covariant derivatives more apparent:
\eq{\spl{
a_{1I}^{\rm old} \, =&\,\, a_{1I}^{\rm new} - c_{IJ} c_1^J\,,\qquad\qquad \tilde a_{1I}^{\rm old} =\, \tilde a_{1I}^{\rm new} - \frac 12 \chi \, c_{1I} - \frac{1}{2} c_{JI}\tilde c_{1}^{\,J} \, , \\
a_{2}^{IJ\,{\rm old}} \, =&\,\, a_{2}^{IJ\,{\rm new}} -\frac{1}{2} \epsilon^{JKL}c_{K}{}^I c_{2L} \, .
}}
Using the new fields the duality relations can be written as
\subeq{\label{DualityFinal}\al{
 \,e^{ 3(4 U + V_1 + V_2 + V_3)} *H_4 \, =&\,\, 2\left[ 2\,c^{IJ}c_{(IJ)} - ({\rm tr}\,c)^2 -2\,\chi \,({\rm tr}\,c) + 3k\right]\, , \label{dualityH4} \\
 \,e^{ 2(4 U + V_1 + V_2 + V_3)}g^{IJ}  *H_{3J} \, =&\,\,  \mathcal Da^I + \epsilon^{IJK} c_{JL}Dc_{K}{}^L   \, , \label{DualityFinalscal}\\
\,e^{4U-V_1-V_2-V_3}g_{IJ} *\tilde H_{2}^J \, =&\, \,  \mathcal Da_{1I}  - 2\,c_{IJ} \,\mathcal Dc_1^J - (c_{IK}c_J{}^K + \epsilon_{IJK} a^K ) F_{2}^J \, , \label{DualityFinalvec1}\\
\, e^{V_1+V_2+V_3} *H_{2I} \, =&\,\, \mathcal D\tilde a_{1I} - c_{JI} \tilde H_2^J - \chi \,\mathcal D c_{1I} \, , \label{DualityFinalvec2}
}}
where we defined the curly covariant derivatives:
\eq{\label{curlycovder}\spl{
\mathcal Da^I \, =&\,\,  Da^I + 2a_{1}^I - 4\tilde a_{1}^I - 6k A^I + 4\, c^{JI}(c_{1J}+\tilde c_{1J}) - 4({\rm tr}\, c)(c_{1}^I+ \tilde c_{1}^I) \, ,\\
\mathcal D a_{1I} \, =&\,\, Da_{1I}+\epsilon_{IJK}\big(-2 c_1^J \wedge c_1^K -2 c_1^J \wedge \tilde c_1^K + 3k A^J\wedge A^K +4 \, a_2^{JK}\big) \, , \\
\mathcal D\tilde a_{1I} \, =&\,\, D\tilde a_{1I} + \epsilon_{IJK}\left(\tilde c_{1}^J\wedge c_1^K + \tilde c_{1}^J\wedge\tilde c_1^K + 2\,a_{2}^{JK} \right)\, , \\
\mathcal D c_{1I} \, =&\,\, Dc_{1I} + 2 c_{2I}\, , \\
\mathcal D \tilde c_{1I} \, =&\,\, D\tilde c_{1I} - 2 c_{2I} \, .
}}
The last one is included because it will be useful in section \ref{sec:N4}. As we will see there, these are the proper covariant derivatives of the full gauge group.
So the scalars $a^I$ are dual to the two-forms $c_{2I}$, while the vectors $a_{1I}$ and $\tilde a_{1I}$ are dual to $\tilde c_{1}^{\,I}$ and $c_{1I}$ respectively. Moreover, though we have not reported the explicit relation (since it is only needed to check that (\ref{EoMtildec1}) is a Bianchi identity for $a_1$), the two-forms $a_{2}^{IJ}$ are dual to the scalars $c_{IJ}$. It will turn out that in our symplectic frame the electric vectors are $A_1^I$, $c_{1I}$ and $a_{1I}$, while the magnetic ones are $\tilde c_{1}^{\,I}$, $\tilde a_{1I}$ (together with the duals of $A_1^I$, let us dub them $\tilde A_{1I}$, which do not appear in our setup, and whose M-theory origin is to be searched in a ``magnetic dual'' of the 7D metric).

\subsection{Equations of motion in the final electric-magnetic frame}
\label{app:eomSWframe}

In the following we perform the field transformations needed to make contact with the general $\mathcal{N}=4$
action of \cite{SchonWeidner}. Concretely, as shown in table \ref{fluxfield} we need to dualize $c_3,c_{2I},\tilde{c}_{1}^{\,I}$ into $k,a^I,a_{1I}$ respectively.

The Bianchi identities for the $c$-fields correspond to the second-order equations of motion for the dual $a$-fields. Conversely, the second-order equations of motion for the $c$-fields are
trivially solved in terms of the $a$-fields and thus correspond to Bianchi identities for the $a$-fields.
Given the system of relations derived in the subsections above, we can define second-order equations of motion for a chosen complete set of independent degrees of freedom, which are appropriate for comparing with the $\mathcal N=4$ formalism.

First we notice that \eqref{dualityH4} can be used to completely eliminate $H_4$ (and therefore the three-form potential $c_3$) from the equations of motion in favor of the flux constant $k$. This is equivalent to solving the total derivative eq.~(\ref{eqH4}).

Second, we observe that eq.~(\ref{EoMc2}) describes propagating two-forms $c_{2I}$, which are not allowed in the formalism of \cite{SchonWeidner}. Therefore we use the duality relations to interpret it as a Bianchi identity for the covariant derivative of the scalars $a^I$; so it is identically solved. On the other hand substituting $H_{3I}$ and $\tilde H_2^J$ with their duals, the Bianchi identity (\ref{BianchiH}) for $H_{3I}$ becomes our new equation of motion, and reads
\begin{multline}
0 = D\left[e^{ -2(4 U + V_1 + V_2 + V_3)}g_{IJ} * \big(\mathcal D a^J + \epsilon^{JKL} c_{KE}Dc_{L}{}^{E} \big)\right] \\
 -\, e^{-4U+V_1+V_2+V_3}\epsilon_{IJK} g^{JL} F_2^{K} \wedge *\left[  \mathcal Da_{1L}  - 2\,c_{LE} \,\mathcal Dc_1^{E} - (c_{LF}c_{E}{}^{F} + \epsilon_{LEF} a^{F}) F_{2}^{E}  \right]  \label{EoMscalara} \, .
\end{multline}

Moreover, in section \ref{sec:veciden} we show that the $\tilde c_{1}^{\,I}$ should be auxiliary magnetic vectors, while the propagating degrees of freedom should instead be their electric duals $a_{1I}$. So we exchange the Bianchi and the equations of motion of $\tilde c_{1}^{\,I}$ and $a_{1I}$ via the duality relations. While eq.~(\ref{EoMtildec1}) is then identically solved, from (\ref{BianchiH}) we obtain the following equation of motion for $a_{1I}\,$:
\begin{multline}
0 = D \left[ e^{-4U+V_1+V_2+V_3}g^{IJ} * \left(\mathcal Da_{1J}  - 2\,c_{JK} \,\mathcal Dc_1^K  - c_{JL}c_K{}^L F_{2}^K - \epsilon_{JKL} a^L F_{2}^K  \right) \right]
  \\
-\, 2\,e^{- 2(4 U + V_1 + V_2 + V_3)}\delta^{IJ}g_{JK} *\left(\mathcal Da^K + \epsilon^{KLE} c_{LF}Dc_{E}{}^F \right) + \d\chi \wedge F_{2}^I \,.
\label{EoMa1}
\end{multline}

In the end the final set of second-order equations of motion for the new propagating fields are eqs.~(\ref{EoMscalara}), (\ref{EoMa1}) for the scalars $a^{I}$ and the vectors $a_{1I}$, together with
eqs.~(\ref{EoMc1}), (\ref{EoMtildec}), and (\ref{EoMcIJ}) for $c_{1I}$, $\chi$ and $c_{IJ}$ respectively. In the latter equations,
we use the duality relations to eliminate $H_4$, $H_{3I}$ and $\tilde H_{2I}$. Doing so, the equation for $c_{1I}$ becomes
\eq{\spl{
& 0 = D\left[e^{V_1+V_2+V_3} *H_{2I}\right] \,+\,  2\,e^{-4U} \epsilon_{IJK} \, g^{KL}* H_{1L}{}^J \, +\, \d\chi \wedge H_{2I} \\
& -\, e^{-4U+V_1+V_2+V_3}  H_{1JI}\wedge *g^{JK}\left[\mathcal Da_{1K}  - 2\,c_{KL} \,\mathcal Dc_1^L - (c_{KE}c_L{}^E + \epsilon_{KLE} a^E ) F_{2}^L\right] \\
& \,+\, 2\, e^{ -2(4 U + V_1 + V_2 + V_3)}\left[ (\chi + {\rm tr}\,c)\delta_{IJ} -2\,c_{(IJ)} \,\right]\delta^{JK}g_{KL}*\left(\mathcal Da^L + \epsilon^{LEF} c_{EG}Dc_{F}{}^G \right)  \, ,
\label{EoMc1Bis}
}}
while the one for $\chi$ reads
\eq{\spl{
& 0 = \d \!\left[e^{-2V_1-2V_2-2V_3}*\d \chi \right] \,+\, \mathcal Da_{1I} \wedge F_2^I   + \mathcal D c_{1I}\wedge \mathcal Dc_1^I  \\
& +\,8\,e^{-8U-3V_1-3V_2-3V_3}g_{IJ}\left[ (\chi + {\rm tr}\,c)\delta^{IJ} -2\,c^{(IJ)} \,\right]*1  \\
& -\, 8 \,e^{ -3(4 U + V_1 + V_2 + V_3)} ({\rm tr}\,c)\left[ 2\,c^{IJ}c_{(IJ)} - ({\rm tr}\,c)^2 -2\,\chi \,({\rm tr}\,c) + 3k\right]*1 \, , \label{EoMtildecBis}
}}
and the one for $c_{IJ}$ is
\eq{\spl{
& 0 = D\left[e^{-4U} g^{IK}\delta^{JL} * H_{1KL}\right] + e^{V_1+V_2+V_3} F_{2}^I \wedge * H_{2}^J + 8\,e^{-12U-V_1-V_2-V_3}({\rm tr}\, c)\delta^{IJ} *1 \\
&+\, 4\,e^{-8U-3V_1-3V_2-3V_3} ( \delta^{IJ} \delta^L_K -\delta^{I}_K \delta^{JL} -\delta^{J}_K \delta^{IL} ) g_{LE}\!\left[ (\chi + {\rm tr}\,c)\delta^{KE} -2\,c^{(KE)} \,\right] *1 \\
&- \,e^{-4U+V_1+V_2+V_3}g^{IK} H_{2}^J \wedge *\left[\mathcal Da_{1K}  - 2\,c_{KL} \,\mathcal Dc_1^L - (c_{KE}c_L{}^E + \epsilon_{KLE} a^E ) F_{2}^L\right]  \\
&+ \,e^{ -2(4 U + V_1 + V_2 + V_3)} \epsilon^{IKE} g_{KL} H_{1E}{}^J \wedge *\left(\mathcal Da^L + \epsilon^{LEG} c_{EF}Dc_{G}{}^F \right) \\
&-4 \,e^{ -3(4 U + V_1 + V_2 + V_3)} \left[ (\chi + {\rm tr}\,c)\delta^{IJ} -2c^{(IJ)} \,\right]\! \left[ 2c^{KL}c_{(KL)} - ({\rm tr}\,c)^2 -2\chi ({\rm tr}\,c) + 3k\right]*\!1 .
\label{EoMcIJbis}
}}
We can now construct the form sector of the 4D action by requiring that it produces precisely the equations listed above.
The result is summarized in section \ref{sec:resultaction}.

\section{Details on the identification of $\mathcal N=4$ fields through E$_{7(7)}$}
\label{app:detailsE77}

In this appendix we study in more detail how the maximal global symmetry E$_{7(7)}$ breaks in the case of tri-Sasakian manifolds
and how this helps in identifying the coset manifold of scalar fields as well as the symplectic frame for the vector fields.

We have seen in section \ref{sec:cosetiden} that it is convenient to consider the subgroup SL(8,$\mathbb{R}$) of E$_{7(7)}$, which
in turn contains the GL(7,$\mathbb{R}$) describing diffeomorphisms of the 7D internal manifold. Concretely, the latter embedding
is as follows:
\eq{
\label{SL8Rrep}
P^a{}_b = \left( \begin{array}{cc} (\det F)^{-1/4} F^m{}_n & \mathbf{0}\\ \mathbf{0} & (\det F)^{3/4} \end{array} \right) \, ,
}
where $P\in  \text{SL}(8,\mathbb{R})$ and $F\in \text{GL}(7,\mathbb{R})$, so the range of the indices is $a,b=1,\ldots, 8$ and $m,n=1,\ldots, 7$. The diffeomorphism $F$ generates
a generic metric by acting in the standard way on the covariant indices of the metric $g_0$ at the origin of the scalar manifold:
\eq{
\label{metricaction}
g_{mn} = (F^{-1})^p{}_m \, g_{0,pq} \, (F^{-1})^q{}_n \, .
}
In our case, the metric $g_0$ we start from is the canonical tri-Sasakian metric \eqref{3Smetric}, while $g$ is a generic 7D metric within the class of our truncation ansatz (\ref{metricansatz})
described by the scalar fields $U$ and $g_{IJ}$. From section \ref{sec:cosetiden} we know already that our truncation ansatz breaks GL(7,$\mathbb{R}$) to SO$(4)\times$GL$(3,\mathbb{R})\times \mathbb{R}_0$. At the same time the four-form representation, which as we saw in eq.~\eqref{133split}, makes up the remainder of (the algebra of) E$_{7(7)}$ breaks as
\eq{\label{4formdecompApp}\spl{
\bf{70} & \rightarrow (\bf{3},\bf{6})_0 \oplus (\bf{1},\bf{1})_+ \oplus (\bf{1},\bf{1})_- \, ,\\
\mu_{abcd} & \rightarrow (\,\mu_{\tilde{I}AB},\,\mu^{\tilde{I}}{}_{\tilde{I}}\,,\,\mu_{ABCD}) \, ,
}}
where the subscripts denote the weight under the action of $\mathbb{R}_0 =\text{SO}(1,1)$. Furthermore, $A,B=1,\ldots,4$ and $\tilde{I},\tilde{J}=1,2,3$ indicate indices in the fundamental of SL(4,$\mathbb{R}$) and SO(3) respectively. Contrary to the SO(3)$_R$ we introduced before, this SO(3) acts only on $B_{\rm QK}$ and not on the fibers. Note that a $\tilde{I}$ corresponds to an insertion of $J^{\tilde{I}}$ in $\mu_{abcd}$, so that it counts for a two-form within the four-form. In this appendix we will reserve $I,J=1,2,3$ (without tilde) for the indices of the fibers acted upon by GL(3,$\mathbb{R}$).

Coming back to the diffeomorphism $F^m{}_n$, we find concretely that it decomposes under the symmetry breaking as
\eq{
F = \left( \begin{array}{cc} \lambda \, \Gamma_4(O) & \bf{0} \\ \bf{0} & N  \end{array} \right) \, ,
}
where the $3\times 3$ submatrix $N^I{}_J \in \text{GL}(3,\mathbb{R})$ transforms the fibers, $O \in \text{SO(3)}$ with $\Gamma_4$ indicating the 4D representation (so $\Gamma_4(O)$ is a $4\times 4$ matrix), and $\lambda \in \mathbb{R}_0$ generates an overall scaling of $B_{\text{QK}}$. Acting on the two-forms $J^{\tilde{I}}$, $O$ will just rotate them as expected.
Plugging this into \eqref{SL8Rrep} we find for $P \in \text{SL}(8,\mathbb{R})$:
\eq{
P = \left( \begin{array}{ccc} (\det N)^{-1/4}\Gamma_4(O) & \bf{0} & \bf{0} \\
\bf{0} & \lambda^{-1} (\det N)^{-1/4} N & \bf{0} \\ \bf{0} & \bf{0} & \lambda^3 (\det N)^{3/4} \end{array} \right) \, .
}
Let us now separate the $\mathbb{R}_0$ part of \eqref{decompSL8} from the rest to find out which scalar
field from the metric sector ends up in SL(2,$\mathbb{R}$). We find that we should split $P$ as
\eq{\label{SL8rep}\spl{
&P = \\&\left( \begin{array}{ccc} \Gamma_4(O) & \bf{0} & \bf{0} \\
\bf{0} & \lambda^{-1} (\det N)^{-1/2} N & \bf{0} \\ \bf{0} & \bf{0} & \lambda^3 (\det N)^{1/2} \end{array} \right)
\left( \begin{array}{ccc} (\det N)^{-1/4} \bbone_4 & \bf{0} & \bf{0} \\ \bf{0} & (\det N)^{1/4} \bbone_3 & \bf{0} \\ \bf{0} & \bf{0} & (\det N)^{1/4} \end{array}
\right).
}}
Indeed the first factor leaves $\mu^{\tilde{I}}{}_{\tilde{I}}$ and $\mu_{ABCD}$ invariant as according to
\eqref{4formdecompApp} should be the case for the action of an element of $\text{SO}(3)\times\text{SL}(4,\mathbb{R})$,
while the second factor leaves $\mu_{\tilde{I}AB}$ invariant as expected from the action of $\mathbb{R}_0$.

We are now ready to construct the complete SL(2,$\mathbb{R}$) factor. From \eqref{metricaction}
we read off that the group element of $\mathbb{R}_0$ that generates the metric ansatz \eqref{metricansatz} from the
standard metric \eqref{3Smetric} has $\det N = \exp (-V_1-V_2-V_3)$, so we see that the action of $\mathbb{R}_0$ corresponds
to an overall rescaling of the metric on the leaves. Furthermore, from \eqref{C3rep} and \eqref{C3pot} we find
\eq{
\frac{1}{4!}\epsilon^{ABCD}\mu_{ABCD}=\frac{1}{3!}\epsilon^{IJK}\mu_{IJK8} = \frac 12 \, \chi  \, .
}
Since it has four covariant indices it will transform under $N$ with a factor $(\det N)^{-1}=\exp(V_1+V_2+V_3)$. \sloppy In the end
we find that the appropriate representative of SL(2,$\mathbb{R}$)/SO(2) is
\eq{
L_{\text{SL}(2,\mathbb{R})} = \left(\begin{array}{cc} 1 & \chi  \\ 0 & 1 \end{array} \right)
.\left(\begin{array}{cc} e^{\frac{V_1+V_2+V_3}{2}} & 0 \\ 0 & e^{-\frac{V_1+V_2+V_3}{2}} \end{array}\right) \, ,
}
Alternatively we find from
\eq{
M_{\text{SL}(2,\mathbb{R})}= L_{\text{SL}(2,\mathbb{R})} L_{\text{SL}(2,\mathbb{R})}^T = \frac{1}{\Im \tau}\left(\begin{array}{cc} |\tau|^2 & \Re \tau \\ \Re \tau & 1 \end{array} \right) \, ,
}
that the SL(2,$\mathbb{R}$)/SO(2) is described by the complex scalar
\eq{
\tau=\chi + i \,e^{V_1 +V_2 +V_3} \, .
}

Next we will analyse the first matrix in \eqref{SL8rep} in order to construct the SO(6,3)-factor of the $\mathcal N=4$ global symmetry group and identify
the scalar fields therein.
It will be convenient to work at the level of the algebra first and consider infinitesimal transformations.
Let us drop the SO(3)-sector, i.e.\ the first three rows and columns, for the moment, and consider
an element of the full sl(4,$\mathbb{R}$) algebra:
\eq{\label{sl4rep}
\mu^A{}_B = \left(\begin{array}{cc} n^I{}_J - (\frac{1}{2} \text{Tr}\, n + \lambda) \delta^I{}_J & \mu^8{}_J\\
\mu^I{}_8 & 3 \lambda + \frac{1}{2} \text{Tr}\, n \end{array} \right) \in \text{sl}(4,\mathbb{R})\, ,
}
where $n \in \text{gl}(3,\mathbb{R})$ is the infinitesimal diffeomorphism corresponding to $N$, $\lambda$ is the infinitesimal scaling, and
we introduced $\mu^8{}_J$ and $\mu^I{}_8$ to fill the off-diagonal blocks. We know already from eq.~\eqref{C6rep} that $\mu^m{}_8$ corresponds
to the dual form potential $\tilde{C}_6$ and thus to the scalar field $a^I$ therein, while we will not need $\mu^8{}_m$ for our choice of gauge for the coset representative.
Now there exists an isomorphism between sl(4,$\mathbb{R}$) and so(3,3), under which the fundamental $\bf{4}$ representation
of the former is the spinor representation of the latter. Likewise, the fundamental $\bf{6}$ representation of the latter is
the antisymmetric product of two $\bf{4}$'s of the former. One can explicitly construct the isomorphism using a convenient representation
of the gamma-matrices of so(3,3). We find that $\mu^A{}_B$ is equivalent to
\eq{
\mu = \left( \begin{array}{cc} n + 2 \lambda \bbone_3 & b_1 \\ b_2 & -n^T - 2 \lambda \bbone_3 \end{array} \right) \in \text{so}(3,3) \, ,
}
with $b_1$ and $b_2$ antisymmetric matrices given by:
\subeq{\label{b1b2so33}\al{
& b_{2\,IJ} \,=\,  \epsilon_{IJK} \mu^K{}_8 \, , \\
& b_1{}^{IJ} \,=\, -\epsilon^{IJK} \mu^8{}_K \, ,
}}
and where we used a ``generalized geometry style'' metric for so(3,3):
\eq{
\eta_{\text{so(3,3)}} = \left(\begin{array}{cc} \bf{0} & \bbone_3 \\ \bbone_3 & \bf{0} \end{array} \right) \, ,
}
which is a submatrix of the full so(6,3) metric $\eta$ given in eq.~\eqref{choiceeta}.

Eq.~\eqref{4formdecompApp} tells us that in order to obtain the full so(6,3) we just put the $\mu_{\tilde{I}AB}$ in the off-diagonal blocks:
\eq{
\label{genmu}
\mu_{\text{so(6,3)}} = \left( \begin{array}{ccc} \rule{1em}{0em} o \rule{1em}{0em}  & \mu_2^T & \mu_1^T\\
\mu_1 & n + 2 \lambda \bbone_3 & b_1 \\
\mu_2 & b_2 & -n^T - 2 \lambda \bbone_3 \end{array} \right) \, ,
}
with $o \in \text{so}(3)$ and
\subeq{\al{
&\mu_{1\tilde{J}}^{\,I} \,=\, k_1 \, \epsilon^{IKL} \mu_{KL\tilde{J}} \, , \\
&\mu_{2\,I\tilde{J}} \,=\, k_2 \, \mu_{I8\tilde{J}} \, ,
}}
where $k_1,k_2$ are some proportionality constants, which could be (partly) determined from a more careful study
of the embedding of SO(6,3) in E$_{7(7)}$. We will fix these constants in such a way that the coset manifold gives the correct kinetic terms.
Comparing with \eqref{C3rep} we see that $\mu_1$ corresponds to the form potential $C_3$, and in particular generates the scalars $c_{IJ}$, while
we do not need to turn on $\mu_2$.

Putting all the pieces together and using eqs.~\eqref{C3C6rep} and \eqref{b1b2so33} we find that we can generate precisely the SO(6,3)-part of the exceptional generalized metric corresponding to our
reduction ansatz by acting with the coset representative given in eq.~\eqref{repunrotated} on the standard exceptional generalized metric.

\bigskip

Having identified the $\mathcal N=4$ scalar manifold, in the following we characterize the electric and magnetic vector fields. This is done by studying the transformation of the vectors under the decomposition of E$_{7(7)}$ in SL(2,$\mathbb{R}$)$\times$SO(6,3). The charge under $\mathbb{R}_0 = \text{SO}(1,1)\subset \text{SL}(2,\mathbb{R})$ determines whether they are electric or magnetic vectors in the symplectic frame of \cite{SchonWeidner}.

Under E$_{7(7)}$ the vectors transform in the fundamental representation. As above, it is useful to consider the decomposition of the fundamental in representations of the
subgroup SL(8,$\mathbb{R}$):
\eq{\spl{
\bf{56} & \,\rightarrow\, \bf{28} + \bf{28'} \, ,\\
x & \,\rightarrow\, (x^{ab},y_{ab}) \, ,
}}
where $x^{ab}$ and $y_{ab}$ transform as an antisymmetric two-vector and two-form under SL$(8,\mathbb{R})$ respectively (see again~\cite{walexcep} for more
explanation). Under the symmetry breaking (\ref{decompSL8}) we find
\eq{\spl{
\bf{28} & \rightarrow (\bf{3},\bf{1})_- \oplus (\bf{1},\bf{6})_+ \, ,\qquad \qquad \bf{28}'  \rightarrow (\bf{3},\bf{1})_+ \oplus (\bf{1},\bf{6})_- \, , \\
x^{ab} & \rightarrow (x^{\tilde{I}},\,x^{AB}) \, , \,\quad\qquad\qquad\qquad
x'_{ab} \rightarrow (x'_{\tilde{I}},\,x'_{AB}) \, ,
}}
where the plus and minus signs indicate the behavior under the action
of $\text{SO}(1,1)$. According to \cite{walexcep}, $x'_{ab}=(x'_{mn},x'_{m8})$, where
$x'_{mn}$ corresponds to the membrane, which couples to $c_{1I}$ or $\tilde{c}_{1}^{\,I}$, and $x'_{m8}$ corresponds
to the KK-monopole, which couples to the magnetic dual $\tilde A_{1I}$ of the metric vector $A_1^I$. On the other hand
$x^{ab}=(x^{mn},x^{m8})$, where $x^{mn}$ corresponds to the five-brane, which couples to $a_{1I}$ or $\tilde{a}_{1}^{\,I}$, and $x^{m8}$ corresponds
to the momentum, which couples to $A_{1}^{I}$. So applied to our case we find:
\eq{\spl{
& x'_{ab} \,=\, \big(\,x'_{\tilde{I}}= c_{1I},\,\,\, x'_{IJ}=\epsilon_{IJK} \tilde{c}_{1}^{K},\,\,\, x'_{8I}=  \tilde A_{1I}\, \big) \, , \\
& x^{ab} \,=\, \big(\,x^{\tilde{I}}= \tilde{a}_1^I, \,\,\,\, x^{IJ}=\,\epsilon^{IJK} a_{1K},\,\,\,\, x^{8I}= A_{1}^{I} \,\big) \, .
}}
From the transformation behavior under SO(1,1) we see that $c_{1I}$, $a_{1I}$, $A_{1}^{I}$ are electric vectors while $\tilde{c}_{1}^{\,I}$, $\tilde a_{1}^{\,I}$, $\tilde A_{1I}$ are magnetic.


\bibliography{consreduction}
\end{document}